\documentclass[aps,prl,reprint,longbibliography,superscriptaddress]{revtex4-2}
\usepackage{graphicx} % Include figure files
\usepackage{dcolumn} % Align table columns on decimal point
\usepackage{bm} % bold math
\usepackage[colorlinks=true, citecolor=blue, linkcolor=blue, urlcolor=blue]{hyperref}% add hypertext capabilities
\usepackage{xcolor}
\usepackage{amsmath, amssymb, amsfonts}
\usepackage{mathtools}
\usepackage{subfigure}
\usepackage{mathrsfs}
\usepackage{sidecap}
\usepackage{verbatim}

\definecolor{purple1}{rgb}{128,0,128}

\usepackage{bigints}
\newcommand{\bea}{\begin{eqnarray}}
\newcommand{\ea}{\end{eqnarray}}

\definecolor{darkpastelgreen}{rgb}{0.01, 0.75, 0.24}

\usepackage{braket}

\begin{document}

% Use the \preprint command to place your local institutional report
% number in the upper righthand corner of the title page in preprint mode.
% Multiple \preprint commands are allowed.
% Use the 'preprintnumbers' class option to override journal defaults
% to display numbers if necessary
%\preprint{}

%Title of paper
\title{Evaluation of metrology using self-consistent approach in the MCTDH theory}
\title{Self-consistent many-body metrology}
% repeat the \author .. \affiliation  etc. as needed
% \email, \thanks, \homepage, \altaffiliation all apply to the current
% author. Explanatory text should go in the []'s, actual e-mail
% address or url should go in the {}'s for \email and \homepage.
% Please use the appropriate macro foreach each type of information

% \affiliation command applies to all authors since the last
% \affiliation command. The \affiliation command should follow the
% other information
% \affiliation can be followed by \email, \homepage, \thanks as well.
\author{Jae-Gyun Baak}
\author{Uwe R. Fischer}
%\email[]{Your e-mail address}
%\homepage[]{Your web page}
%\thanks{}
%\altaffiliation{}
\affiliation{
 Seoul National University, Department of Physics and Astronomy,\\
 Center for Theoretical Physics, Seoul 08826, Korea
}

%Collaboration name if desired (requires use of superscriptaddress
%option in \documentclass). \noaffiliation is required (may also be
%used with the \author command).
%\collaboration can be followed by \email, \homepage, \thanks as well.
%\collaboration{}
%\noaffiliation

\date{\today}

\begin{abstract}
We investigate performing classical and quantum metrology and parameter estimation by 
using interacting trapped bosons, which we theoretically treat  by a 
self-consistent many-body approach of the multiconfigurational Hartree type.
Focusing on a tilted double-well geometry, we compare a self-consistently determined and monitored 
two-mode truncation, with dynamically changing orbitals, to the conventional two-mode approach 
 of fixed orbitals, where only Fock space coefficients evolve in time.
We demonstrate that, as a consequence, various metrological quantities associated to a 
concrete measurement such as the classical Fisher information and the maximum likelihood estimator are deeply affected by the orbitals' change during the quantum evolution. 
Self-consistency of the quantum many-body dynamics of interacting trapped ultracold gases 
thus fundamentally affects the attainable 
parameter estimation accuracy of a given metrological protocol. 
 \end{abstract}

% insert suggested keywords - APS authors don't need to do this
%\keywords{}

%\maketitle must follow title, authors, abstract, and keywords
\maketitle

% body of paper here - Use proper section commands
% References should be done using the \cite, \ref, and \label commands
% \section{}
% Put \label in argument of \section for cross-referencing
% \section{\label{}}
% \subsection{}
% \subsubsection{}

Within the currently emerging quantum era, quantum metrology \cite{Helstrom1967,Helstrom1969,Braunstein_1994,Braunstein1996,Vittorio2006,Howard2010,Toth_2014,Braun} has proven itself 
 to be a powerful tool  for  the accurate estimation of even very small physical parameters, such as 
 gravitational wave amplitudes \cite{Schnabel}, or to limit the attainable measurement accuracy of 
 fundamental constants like the speed of light \cite{Schneiter}.  As a result, quantum metrology promises to 
 revolutionize %, in the very near term,  
 the existing technologies of measurement. 
 
 While quantum metrology has frequently been employed in the quantum optical context \cite{Vittorio2011,Pirandola,10.1116/5.0007577,PRXQuantum.3.010202}, 
 more recently the corresponding experiments and theory are also exploring coherent matter waves cf., e.g., Refs.~\cite{PhysRevA.72.043612,PhysRevA.77.053613,Gross2010,Jan2010,Juha2012,Gross_2012,Berrada,PhysRevLett.113.103004,Strobel2014,Karol2014,Huang,Stephen2016,Chengyi2017,RevModPhys.90.035005,Czajkowski,Haine}.
 Photons freely propagating in the quantum vacuum 
 are to a very good approximation noninteracting particles and are well described by plane waves
 of definite momentum. 
 Matter waves forming Bose-Einstein condensates at very low temperatures 
 are, however, interacting by the scattering of their elementary atomic or molecular constituents, 
 and are spatially confined (trapped) by arbitrary scalar potentials. %, e.g. harmonic wells or optical lattices).  
 In what follows, we show that 
 %whenever the particles involved in the metrological protocol 
 %are interacting and trapped, as for experimentally realized Bose-Einstein condensates, 
 %for a given measurement 
 the full self-consistency of the quantum many-body evolution of such a system needs in general 
 to be taken into account,  to yield reliable parameter estimation. 
We demonstrate that the interplay of Fock space amplitudes and time-dependent %shape of the 
field operator modes ($\coloneqq$\,orbitals), representing %the hallmark of 
self-consistent many-body evolution, 
{is crucial.  %. {This is 
This interplay is not obtained when fixing the orbitals' shape, thereby significantly restricting
the associated Hilbert space.}
% along the %whole set of 
%Hilbert space trajectories
%accessicol{ble to the system.}   

 {We take as an archetypical model system and for concreteness a tilted double well, 
 where the parameter to be estimated is the linear slope $p_4$  which could, e.g., represent 
 exposing the gas to constant gravitational acceleration  (see Fig.~1).  
 To facilitate comparison with conventional interferometry, 
 we operate %stay for the whole time evolution 
 within a (continuously monitored) two-mode approximation (TMA), 
 corresponding to two interferometric arms. 
We consider a simple 
(Mach-Zehnder type) experiment 
which counts %at the instant of measurement 
the final number of particles on the left and right. 
It is demonstrated that while a non-self-consistent evolution yields a null result for $p_4$ 
[zero classical Fisher information  %$F_{p_4}$ 
(CFI)], a self-consistent quantum many-body evolution gives finite CFI, %thus 
enabling $p_4$ estimation.}
We therefore conclusively 
show that self-consistency is %generally 
crucial for the correct interpretation of parameter estimation data in a trapped interacting many-body system. We note that Heisenberg-limit number scaling of estimation precision with $1/N$ [instead of the shot noise (standard quantum) limit $\propto 1/\sqrt N$], for cat state distributions in Fock space \cite{RevModPhys.90.035005}, or an enhanced $N$-scaling 
relying on  $k$-body interactions \cite{BoixoPRL,Sergio2008}, are not our aim in the present work. The  focus is on the fundamental imprint of the self-consistency of many-body evolution 
on the accuracy of parameter estimation from the metrological protocol.

\begin{figure}[b]
\includegraphics[width=0.42\columnwidth]{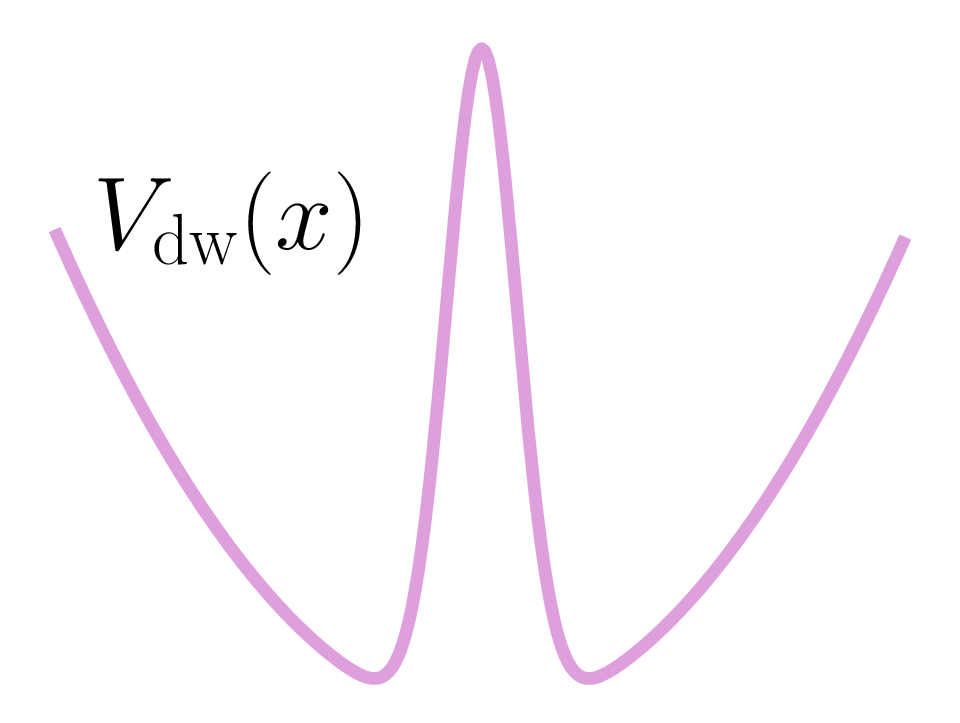}
\includegraphics[width=0.42\columnwidth]{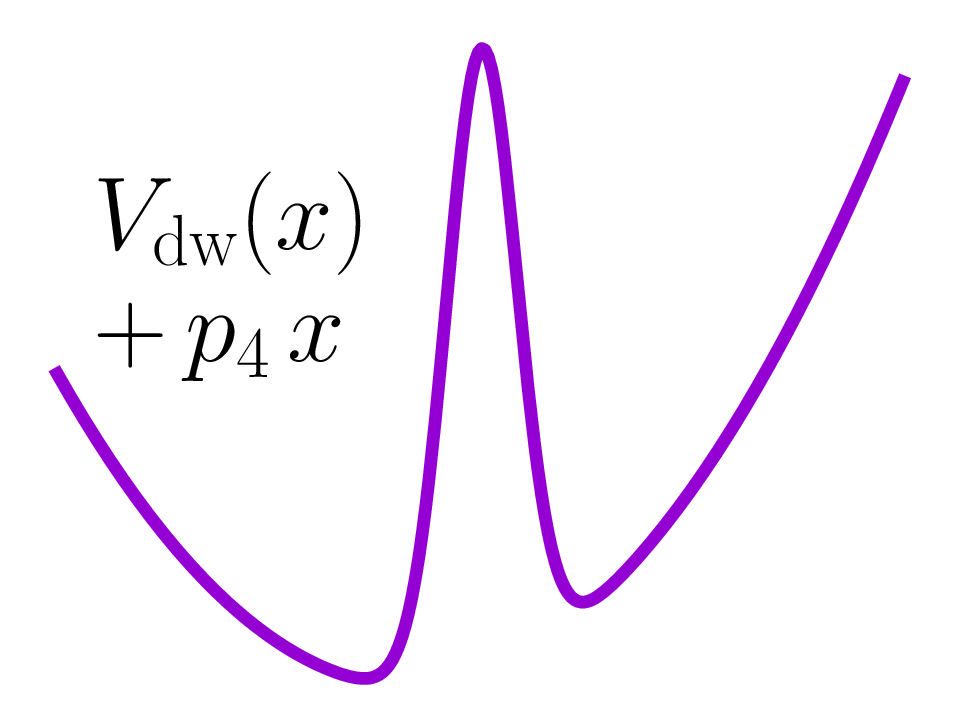}
\includegraphics[width=0.42\columnwidth]{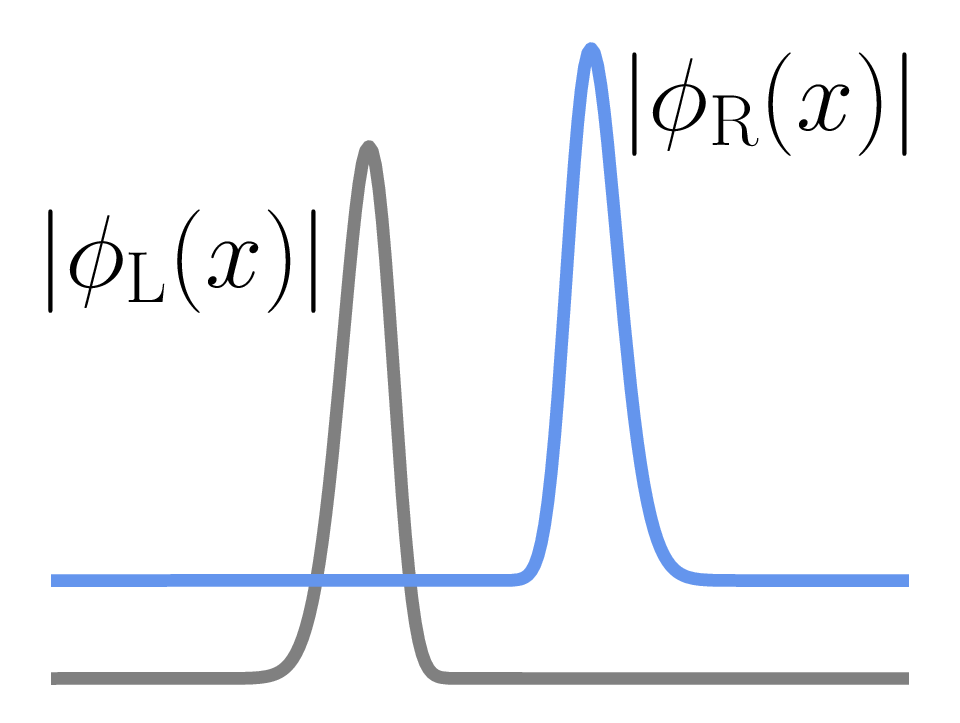}
\includegraphics[width=0.42\columnwidth]{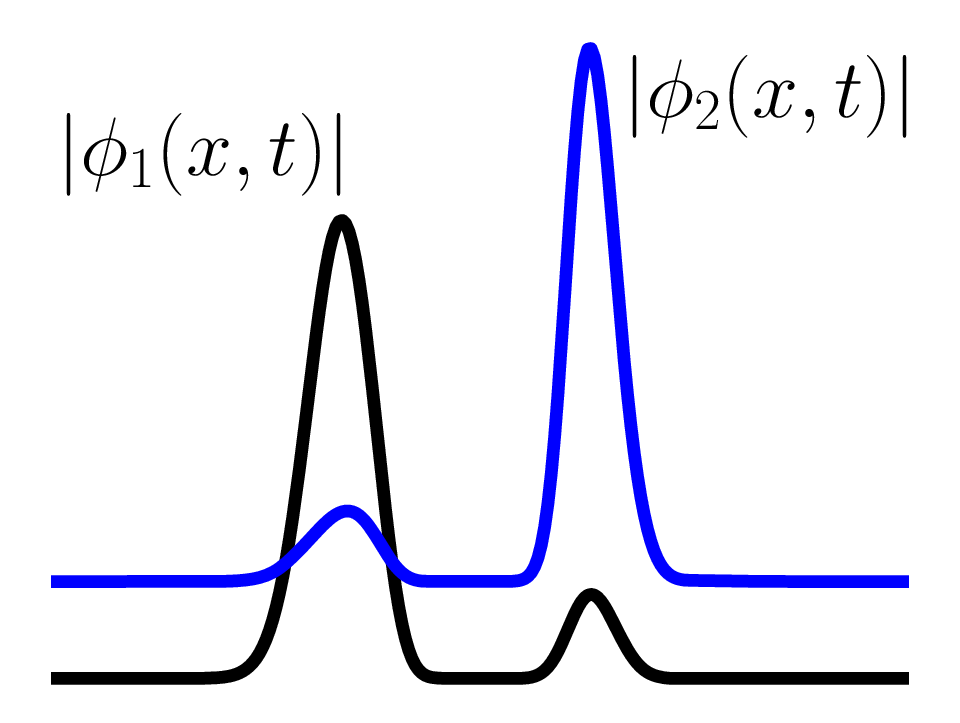}
\caption{\label{fig:orbitals} Top: Trap potential. Left: $t<0$, symmetric double well. Right: 
At $t=0$, the tilt is switched on. Bottom: Initially localized orbitals
%which are used to describe an ultracold atomic gas 
in the large-barrier double-well self-consistently evolve in time %according 
%according to the MCTDH time evolution equations Eq.~\eqref{eq:evolution}.
%Initially, tje symmetric double-well potential  
after tilting, and become delocalized, {whereas without self-consistency
they %conventionally assumed to 
remain localized.}
The orbitals are vertically offset for clarity. 
%Then, after some time, each orbital initially localized on the left ($\phi_{\rm L}$) or the right ($\phi_{\rm R}$) becomes
%delocalized. % loses its localization
%and becomes a different one-particle function: $\phi_1$ or $\phi_2$, respectively.
}
\end{figure}

{To determine the many-body 
evolution self-consistently,  %during the metrological protocol, 
we perform
a multiconfigurational time-dependent Hartree (MCTDH) analysis.
This is a well tested  method to describe the dynamics of spatially confined quantum systems, such as
the one under investigation here, for which self-consistency is crucial} 
% in distinction to translationally invariant systems}  
%which makes use of a time-dependent Fock basis
%and introduces more flexibility to trace the time evolution of the ultracold atomic systems
\cite{Alexej2006,Ofir2008,Axel2020,Rui2020,Ofir2005}.
%In the MCTDH theory,
The general $N$-body state reads %is described by %the ansatz
\vspace*{-0.5em}
\begin{equation}
|\Psi(t)\rangle=\sum\limits_{\vec{n}}C_{\vec{n}}(t)|\vec{n}(t)\rangle\,,
\label{eq:ansatz}
\vspace*{-0.25em}
\end{equation}
where $\sum_{\vec{n}}|C_{\vec{n}}|^2=1$ for state normalization 
and $\vec{n}$ denotes the set of occupation numbers $\{n_i\,|\,i=1,2,\cdots,M\}$  in each mode (orbital), with 
$\sum_{i=1}^M n_i =N$. % the total particle number.
%, with  $Mma$ the total number of orbitals. 
The time-dependent Fock basis state $|\vec{n}(t)\rangle$ indicates that
the orbitals change in time   %according to the time evolution
as a result of finite $M$ (see, Fig.~\ref{fig:orbitals}).
%In this framework, the orbitals as well as the coefficients change as time passes
Their dynamics follows the system of nonlinear coupled integrodifferential 
Eqs.~(S1) in the supplement \cite{suppl}. %, section I. 
Numerical solution enables the %precise 
determination of Fock space coefficients $C_{\vec{n}}$
and orbitals in a self-consistent manner 
at any time \cite{details}.

The quantum metrological approach to parameter estimation, see for example %cf. also, e.g., 
Refs.~\cite{Fujiwara1995,Nielsen2000,PhysRevA.80.032103,Alipour2014,Tomasz2016}, proceeds essentially 
as follows. 
%Basically, it is about enhancing the estimation precision of physical parameters
%by trying to select the efficient initial states, dynamics, measurements, and estimators.
{An initial state $|\psi\rangle$ experiences a dynamical evolution, e.g., $e^{-i\hat{H}_X t}$, during the time $t$
and the final state $|\psi_X\rangle$ contains the information of the parameter $X$.
One chooses an appropriate measurement on $|\psi_X\rangle$ 
to estimate $X$. % based on the measurement results.
%The precision is evaluated by a few kinds of Fisher information.
Previous studies on quantum metrology with ultracold atoms %cf., e.g., Refs.~
\cite{PhysRevA.72.043612,PhysRevA.77.053613,PhysRevA.80.032103,Gross2010,Jan2010,Juha2012,Gross_2012,Berrada,PhysRevLett.113.103004,Strobel2014,Karol2014,Huang,Stephen2016,Chengyi2017,RevModPhys.90.035005,Czajkowski,Haine} 
%e.g. \cite{Juha2012,Karol2014,Huang,RevModPhys.90.035005},  
%\col{put more citations} 
%\col{we should quote here a sizable number of these previous studies} 
have focused on the coefficients $C_{\vec{n}}(X;t)$ in Eq.~\eqref{eq:ansatz},  
%$|\psi_X\rangle=\sum_{\vec{n}}C_{\vec{n}}(X)|\vec{n}\rangle$
and have calculated the quantum Fisher information (QFI) ${\mathfrak F}_{X}$ from $C_{\vec{n}}(X;t)$ only.
However, since the orbitals also evolve by Eqs.~(S1) and the time evolution relies on $X$,
the $|\vec{n}(X;t)\rangle$ must be considered in the calculation of both the QFI and the CFI. }%Fisher information.
%The Fisher information fundamentally implies the distinguishability of states
%when they have slightly different parameter values.
%In order to determine the parameter more exactly,
%the state has to be more sensitive to the change of parameter.
To evaluate the sensitivity of a quantum mechanical state  to a parameter change 
thus requires full exploitation of the information encoded in the state.
Here, we make full use of the parameter dependence of the state, reflected in both coefficients and orbitals.
%and show if there is any discrepancy in the Fisher information between the conventional method and the self-consistent (SC) one,
%where the latter means a method involving in the time-dependent orbitals.
{As a result, we establish numerically exact many-body parameter
estimation for trapped interacting quantum gases.
% and the importance of self-consistency in the estimation process. } %in many-body parameter estimation.} % atomic systems.

%The system we use for quantum metrological protocol is the
%We consider as our model system a
A scalar %single component 
bosonic gas with contact interactions, trapped in a %symmetric 
quasi-one-dimensional (quasi-1D) 
double-well potential,  %, where bosons interact via a $\delta$-function potential 
is described by the Hamiltonian% {($\hbar =m =1$)} 
\begin{equation}
\hat {H}=\sum_{j=1}^N\Big\{-\frac12\frac{\partial^2}{\partial x_j^2}+V(x_j)\Big\}+g\sum^N_{j<k}\delta(x_j-x_k), 
\label{eq:Hamiltonian}
\end{equation}
where $V$ is the trap potential.  
We render $\hat H$ in a dimensionless form, % such that the energy unit is $L^2\coloneqq 1$, where 
fixing a unit length $L$;  the unit of time is then $L^2$ for  $\hbar =m =1$ 
\footnote{In $^{87}\text{Rb}$, $L=1\,\,\mu$m 
%with 24 grid points yields a total size $12\,\,\mu$m and 
yields a time unit $\Delta t =1.366$\,\,msec.}. % For $gN=0.1$, 
%corresponds for 
%$N=10$, %and in quasi-1D,  
%with a transverse confinement length %harmonic oscillator
%$a_\perp \simeq 2 L$
%\,\,\mu$m, %$[\omega_\perp\simeq 10]$kHz, 
%one can render the 3D s-wave scattering length $a_s =a_\perp^2 g_{1D}/4$ 
%away from geometric %scattering 
%resonances \cite{Olshanii1998} %or Feshbach resonances 
%one can  %3D 
%s-wave scattering length $a_s =a_\perp^2 g_{1D}/4 \simeq 0.2$\,\,nm.}
%close to the natural $a_s\simeq 5.77\,\,$nm.} %of  $^{87}\text{Rb}$.}. 
% so that one can use dimensionless position variable $x_j$
%which represents the position of $j$th particle as a proportion of $L$.
%We use $-12\sim 12$ for the range of $x$ in the numerical calculation, which means $24L$ covers the whole size of the system.
%If a BEC is realized with %of $m=1.44\times10^{-25}kg$ ,
The quasi-1D interaction coupling $g$ is % is assumed to be 
controllable by either Feshbach or geometric scattering
resonances %and by transverse trapping 
\cite{Olshanii1998}.  
{Exact solutions of the Schr\"odinger equation % {\em Lieb-Liniger model} encapsulated by  
associated to Eq.~\eqref{eq:Hamiltonian} exist for homogeneous gases under %either 
periodic boundary conditions or in a box trap;   % \cite{Gaudin}. 
their metrology %l properties %ramifications  %of these exact solutions 
was explored in \cite{Jaegyun}.} 

We assume a trap potential of the form $V(x)=V_{\rm dw}(x) + p_4x = \frac{1}{2}p_1x^2+p_2\exp[-x^2/(2\,p_3^2)]+p_4\,x$, where $p_4=0$ initially, cf.~Fig.~\ref{fig:orbitals}.
%The values of t
The remaining parameters, $p_1=0.5$, $p_2=50$, and $p_3=1$, are fixed 
%to keep the validity of the TMA 
throughout the evolution. % we considered.
%These values realize the barrier thick and high enough to make the two single-particle stationary states of the lowest energies degenerate.
The two lowest-energy single-particle states are symmetric and antisymmetric with respect to the origin, 
respectively, 
and their addition and subtraction %of them 
results in two well-localized orbitals:
left $\phi_{\rm L}(x)$ and right $\phi_{\rm R}(x)$, see Fig.~\ref{fig:orbitals}.
These orbitals, approximate ground states of each well,
%are the basic one-particle states that 
furnish the two modes \cite{MilburnCorney1997}. 

\begin{figure}[t]
\includegraphics[width=0.4925\columnwidth]{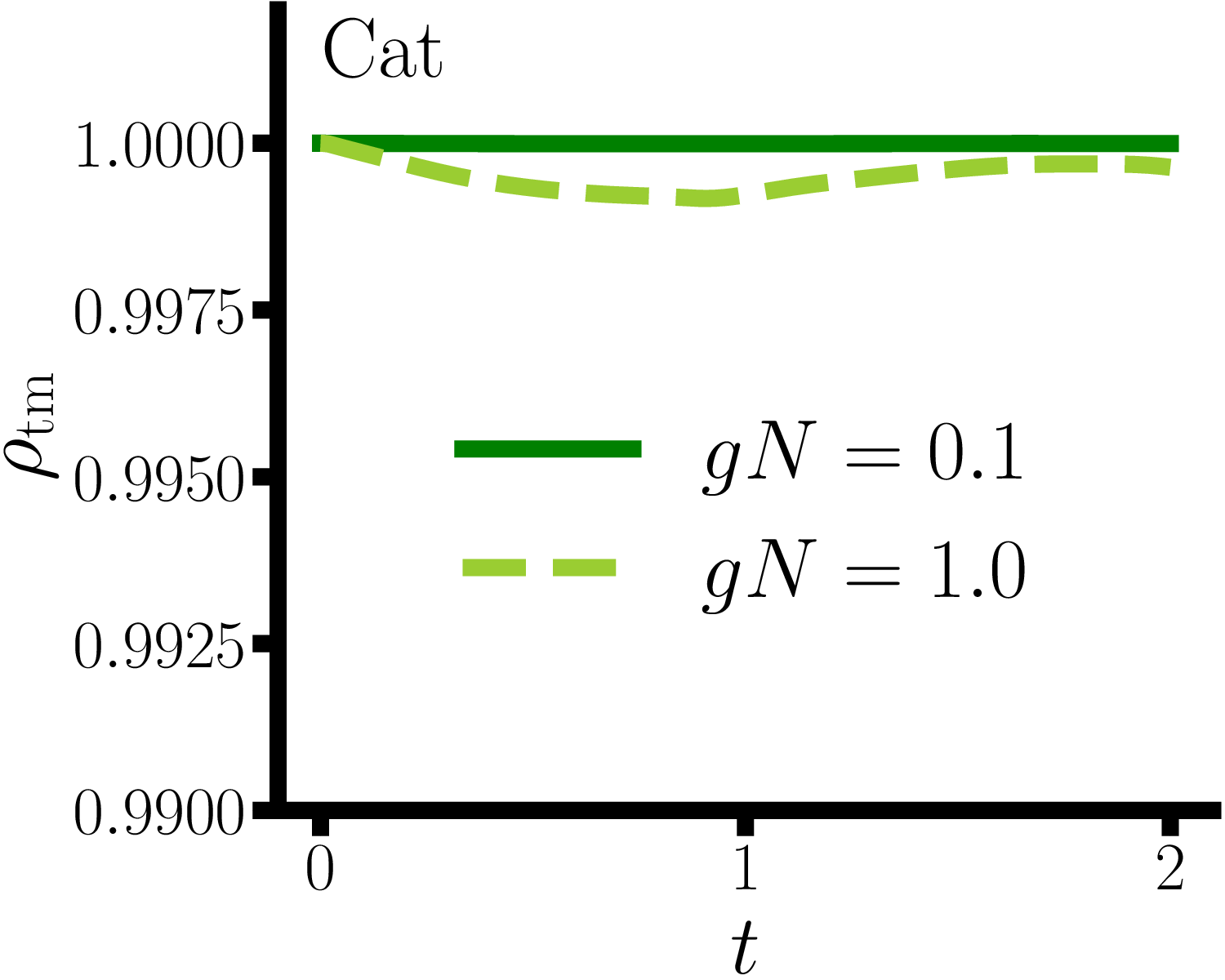}
\includegraphics[width=0.4925\columnwidth]{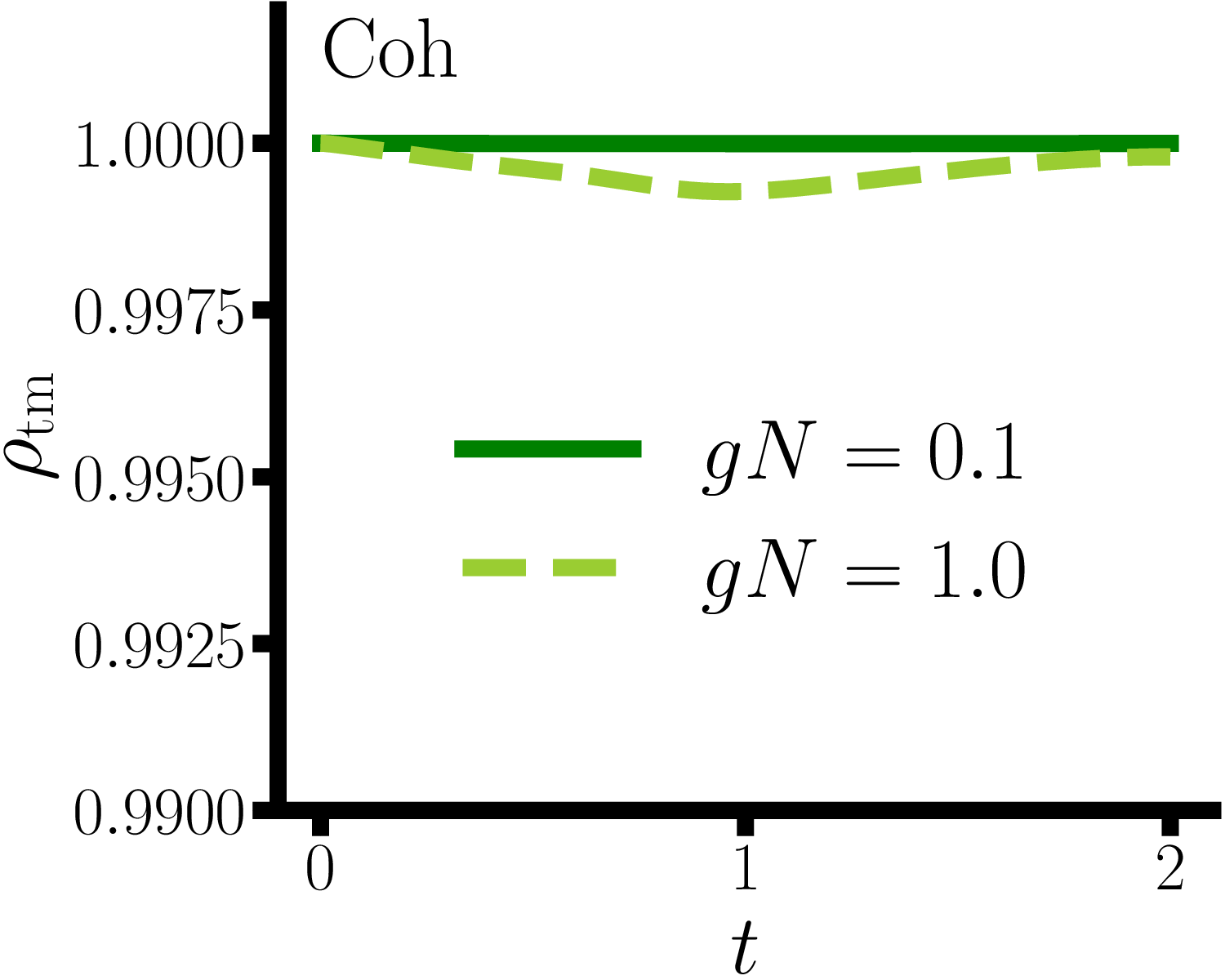}
\caption{\label{fig:two-mode} Monitoring the two-mode truncation after turning on $p_4$,   
verifying whether $\rho_{\rm tm}= (\rho_1+\rho_2)/N\lesssim 1$ (we put $N=10$), for 
%, with  $gN$ values as indicated. 
%, which is the sum of two largest eigenvalues of the reduced one-body density matrix.
cat state (left) and spin-coherent state (right). }  % for the distribution of coefficients.
%Solid and dashed lines are for $gN=0.1$ %, a weakly interacting case,
%and $gN=1$, respectively.} %, an intermediate coupling.}
\end{figure}

The estimation process of $p_4$ proceeds with 
an initial state in the form of Eq.~\eqref{eq:ansatz}. % is supposed to be created.
With $\phi_{\rm L}(x)$ and $\phi_{\rm R}(x)$, we employ two coefficient distributions: 
A NOON (cat) state, %which has
$|\psi_0\rangle=(|N,0\rangle+ %e^{i\alpha}
|0,N\rangle)/\sqrt{2}$, and a spin-coherent state
$|\psi_0\rangle=\sum_{k=0}^N\sqrt{\frac{N!}{k!\,(N-k)!}}
%\cos^{N-k}(\frac{\theta}{2}) %\,%e^{ik\phi}
%\sin^k(\frac{\theta}{2})
\cos^{N-k}(\frac{\pi}{4}) %\,%e^{ik\phi}
\sin^k(\frac{\pi}{4})
|N-k,k\rangle$.  
%\col{we should  write them explicitly already here} 
Then a small $p_4$ tilt %, which was initially $0$, 
%of the trap 
is switched on. %,  tilting trap potential and making it asymmetric. 
The state then evolves according to Eqs.~(S1); finally, 
the number of particles in each well is measured and %the value of 
$p_4$ is estimated from the set of outcomes.   % from the set of measurement outcomes.

%We are using two modes, i.e., $M=2$, for the self-consistent dynamics.
As the relative interaction strength $gN$ (typical ratio
of interaction over single-particle energies), increases, 
more modes than two are required to correctly reproduce the many-body dynamics \cite{Ofir2005}.
Keeping $gN$ fixed when taking the limit $N\rightarrow\infty$ has been demonstrated 
 to reproduce the Gross-Pitaevskii ground state energy \cite{Lieb3D,Lieb1D}.
We thus keep in the following $gN$ fixed when varying $N$, to remain close to the TMA.  
In order to adequately compare self-consistent (SC) evolution to %results to those of the 
conventional SU(2) two-mode interferometry (TMI),
which operates with changing Fock space coefficients only, 
%the SC method has to describe almost exactly the real dynamics of the system.
%To this end,
we maintain TMA validity %of the TMA
throughout the time evolution. %certifying % under nonzero $p_4$.
%its validity 
This can be assessed by evaluating $\rho_{\rm tm}\coloneqq (\rho_1+\rho_2)/N$, cf.~Fig.~\ref{fig:two-mode}, 
where $\rho_j$ is the $j$th largest eigenvalue of the reduced one-body density matrix, 
$\rho^{(1)}(x,x';t)$; see (S2) in the supplement \cite{suppl}. After
diagonalization, $\rho^{(1)}(x,x')=\sum_j\rho_j(t)\,\phi_j^{({\rm no})\ast}(x',t)\,\phi_j^{({\rm no})}(x,t)$, 
with the {\em natural orbitals} $\{\phi_j^{({\rm no})}(x,t)|\,j=1,2,\cdots\}$. 
When a single $\rho_1=O(N)$ (in the formal limit $N\rightarrow\infty$), 
%and the other $\rho_j$ are not $O(N)$, 
a Bose-Einstein condensate is obtained. %, and 
For several %more than one 
$\rho_j$ of $O(N)$, we have a fragmented condensate \cite{Oliver1956}.
The validity of the TMA (two-fold fragmented condensate)  
depends on %the extent to which 
whether $\rho_1\simeq\rho_2\simeq N/2$ and $\rho_{\rm tm}\simeq O(1)$ hold. 
In the supplement \cite{suppl}, we provide further evidence for the appropriateness of using the TMA,  
by demonstrating the rapid convergence of our results with increasing $M$. 
%In such case, the bosonic dynamics mainly relies on those two orbitals, i.e., $\phi_1^{({\rm no})}$ and  $\phi_2^{({\rm no})}$,
%and hence the two-mode description gets valid.

We define an initial state using four orbitals ($M=4$), which are including $\phi_{\rm L}(x)$ and $\phi_{\rm R}(x)$, and then monitor the natural occupations $\rho_j$ under self-consistent evolution at nonzero $p_4$, 
% Also, two different distributions of coefficients for the initial state were tested: 
for both cat and spin-coherent states.
Fig.~\ref{fig:two-mode} shows %the monitoring of 
$\rho_{\rm tm}$, where 
%so how macroscopically the first two orbitals, $\phi_1^{({\rm no})}(x,t)$ and $\phi_2^{({\rm no})}(x,t)$, are occupied by particles is being watched as time goes.
we observe it is close to unity. Also, both  %, the more valid the TMA is.
 $\rho_1$ and $\rho_2$ are macroscopically occupied during the evolution, with negligible occupations 
$\rho_3$ and $\rho_4$. % compared to the formers.
This however also depends on the parameter regime used.
When $gN=0.1$, two modes are sufficient, but when $gN=1$, $\rho_{\rm tm}$ discernibly dips below unity.
%Furthermore, i
Increasing $p_4$ further,  
%is very large, its effect of disturbance makes the particles excited, thus 
%$\rho_{\rm tm}$ increasingly deviates from unity and 
the TMA fails. 
An appropriate regime of parameters where $\rho_{\rm tm}\simeq 1$ %is necessary,
is obtained by fixing $gN=0.1$ and $p_4=0.1$.
Also, even though natural orbitals are used in the discussion above,
we can apply the two-mode criterion to the left/right orbitals or their time-evolved forms, i.e., $\phi_1(x,t)$ and $\phi_2(x,t)$; there always exists a unitary transformation such that $\phi_j(x,t)=\sum_{jk}U_{jk}\,\phi_k^{({\rm no})}(x,t)$. 
%keeping the physics of the system constant.
%\col{When two natural orbitals are sufficient, this guarantees that the number of
%other unitarily transformed orbitals is two as well.} 
%In conclusion, $\phi_1(x,t)$ and $\phi_2(x,t)$ are sufficient for the exact description of the system throughout the time evolution when $gN=0.1$ and $p_4=0.1$.

Expanding the second-quantized form of Eq.~\eqref{eq:Hamiltonian} 
with $\hat{\Psi}(x)=\hat{b}_{\rm L}\,\phi_{\rm L}(x)+\hat{b}_{\rm R}\,\phi_{\rm R}(x)$,
a two-site single-band Bose-Hubbard Hamiltonian is obtained. %, describing the bosons in a double well. 
In terms of the usual SU(2) Pauli matrices %spin operators 
$\hat J_x = \frac12 (\hat b_{\rm L} \hat b_{\rm R} + \hat b^\dagger_{\rm R} \hat b_{\rm L})$ and 
$\hat J_z =  \frac12 (\hat b_{\rm L} \hat b_{\rm L}- \hat b^\dagger_{\rm R} \hat b_{\rm R})$, 
%\col{write here which form the spin operators take and put citation} 
%which describes the quantum dynamics of a two-mode systems, i.e., the bosons in a double-well potential:
\begin{equation}
\hat{H}=-\tau\hat{J}_x+\epsilon\,\hat{J}_z+U\hat{J}_z^2\,,
\label{eq:bose-hubbard}
\end{equation}
where $\tau$ is tunneling amplitude, $\epsilon$ an energy offset between wells,
and $U\propto g$ an interaction coupling, all depending on integrals involving the two orbitals.  
%\cite{Ofir2008}.} % \col{related how to $g$?}.
%The other parameters in front of the terms such as $\hat{J}_x\hat{J}_y$ or $\hat{J}_y^2$ are extremely small
%compared to the parameters in Eq.~\eqref{eq:bose-hubbard} due to the geometry of the system.
%Previous studies, e.g.,  
%\col{again, quote some of these studies here} 
The implementation of quantum metrological protocols 
using the above Hamiltonian was carried out, e.g., in Refs.~\cite{Juha2012,Karol2014,Tomasz2016}:  % \col{citations},
An initial state $|\psi_0\rangle$, %therein completely 
as defined by the distribution of coefficients $C_{\vec n}$,  
evolves as $\exp(-i\hat{H}t)|\psi_0\rangle$, 
and the parameter of interest, e.g., $\epsilon$, is estimated from the population imbalance 
%between the two modes
\cite{Juha2012,Karol2014}.
In our setup, the strong barrier renders the {initial $\tau$ exponentially small} compared to $\epsilon$ and $U$,
and $\epsilon\neq0$, as the symmetry of $V(x)$ is broken by $p_4$.
Hence the {TMI} time evolution operator is, {to very good accuracy,}  $\exp(-i(\epsilon\hat{J}_z+U\hat{J}_z^2)t)$,
and the QFI can be analytically calculated by 
$\mathfrak{F}_\epsilon=4\,\langle\psi_0|(\Delta\hat{J}_z)^2|\psi_0\rangle$.
When the cat state %$|\psi_0\rangle=(|N,0\rangle+e^{i\alpha}|0,N\rangle)/\sqrt{2}$ 
is used, $\mathfrak{F}_\epsilon=N^2\,t^2$, which is denoted as the Heisenberg limit.
For the spin-coherent state, 
%$|\psi_0\rangle=\sum_{k=0}^N\sqrt{\frac{N!}{k!\,(N-k)!}}\cos^{N-k}(\frac{\theta}{2})\,e^{ik\phi}\sin^k(\frac{\theta}{2})|N-k,k\rangle$,
$\mathfrak{F}_\epsilon=N\sin^2(\theta)\,t^2$, representing the shot-noise (standard quantum) limit.
Any nonzero $\tau$ deteriorates the $N$-scaling of $\mathfrak{F}_\epsilon$, 
confirmed by numerically calculating the QFI \cite{Juha2012}.
Note that here only the change of Fock space coefficients has been considered,
while the orbitals are fixed in TMI. %supposed to remain unchanged. % during the dynamical evolution.
Because of the latter fact, the still exponentially small 
$\tau$ and $U$ are kept constant during the evolution,
and $\epsilon$ is abruptly switched on at $t=0$. %, by turning on $p_4$. % of the potential.
{In our setting, $\epsilon\in (-0.7,-0.6)$, and $U\in(0.002,0.03)$, with concrete values 
determined by  %contingent on %are functions of %\col{exact values to achieve what?} 
$gN$ and $N$, on which in turn the initial orbitals $\phi_{\rm L}(x)$ and $\phi_{\rm R}(x)$ depend.  }
Recall that our target parameter is $p_4$, not $\epsilon$,
thus by using the chain rule, $\mathfrak{F}_{p_4}=\mathfrak{F}_\epsilon\times(h_{1}-h_{2})^2$,
where the single-particle energies 
$h_{i}\coloneqq \int dx\,\phi_i^\ast(x) \left[-\frac12\frac{\partial^2}{\partial x^2}+V(x)\right] \phi_i(x)$. 

First, we compare the QFIs of the SC approach and the conventional TMI.
% employing the Bose-Hubbard model with fixed orbitals.
The QFI with respect to a given parameter $X$ inscribed onto a pure state $|\psi_X\rangle$, is
$\mathfrak{F}_X=4\big(\langle\partial_X\psi_X|\partial_X\psi_X\rangle
-|\langle\psi_X|\partial_X\psi_X\rangle|^2\big)$, 
and insertion of Eq.~\eqref{eq:ansatz} into $|\psi_X\rangle$ gives Eq.~(S4),
which facilitates calculation of the QFI using the ingredients of MCTDH from the expansion in Eq.~\eqref{eq:ansatz}, 
{also cf.~%section II in 
\cite{suppl}.} 
The first row in Fig.~\ref{fig:qfi} shows QFI versus time $t$ and particle number $N$, respectively.
For each initial state, the SC method reproduces very well the QFI predicted by TMI.
The influence of self-consistency thus plays a subdominant role for $\mathfrak{F}_X$.
The latter completely depends on the final state,  
and self-consistency (changing orbitals) essentially 
represents fitting that state more exactly.
When the bosons weakly interact ($gN=0.1$) and the disturbance to the system is small ($p_4=0.1$), 
conventional TMI therefore approximates well the %is thus appropriate to capture the 
QFI \cite{explanation}.
%, leading to 
%the combination of both effects leads to the 
%QFI values close to those predicted by %conventional 
%TMI quantum metrology.

\begin{figure}[t]
\includegraphics[width=0.4925\columnwidth]{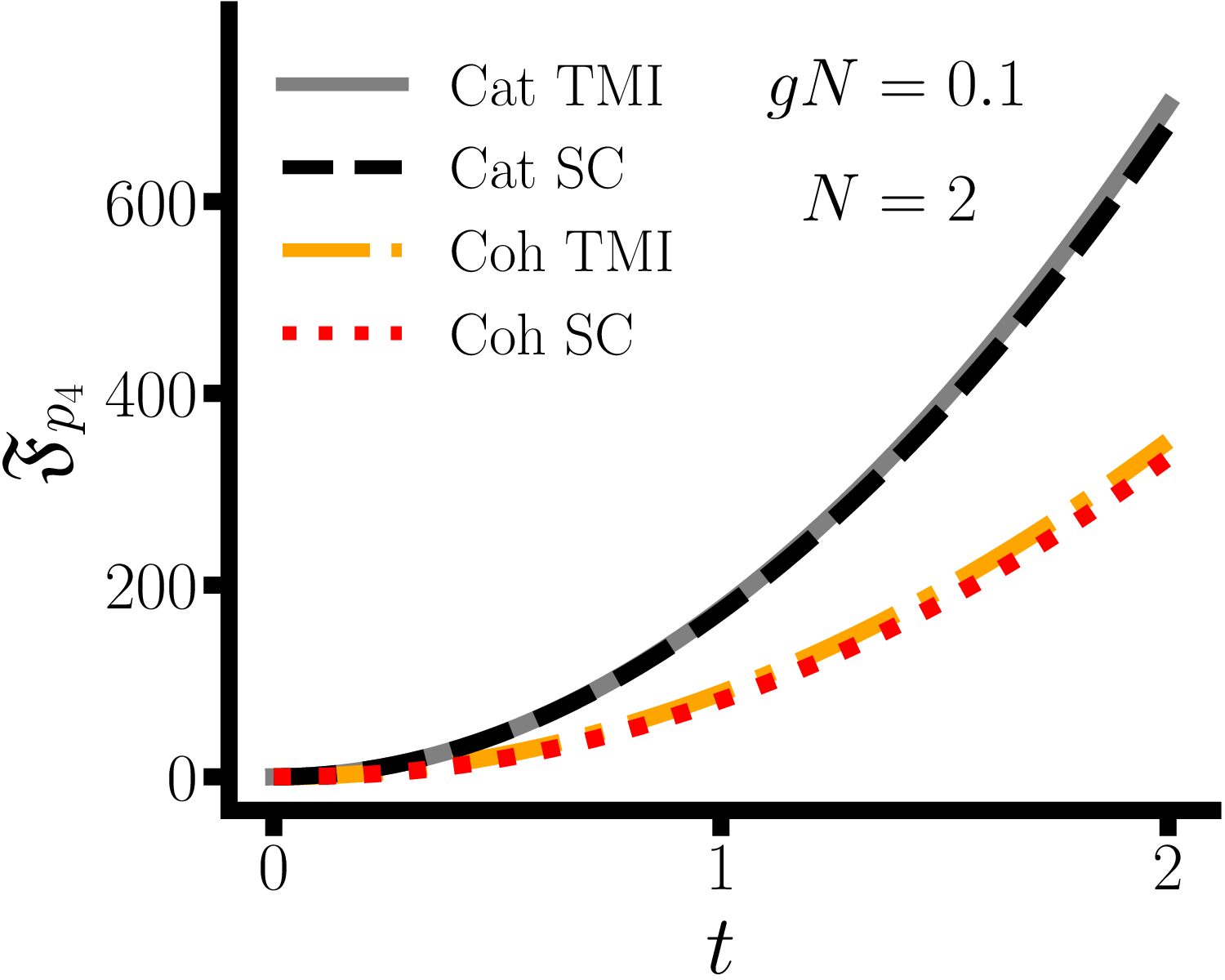}
\includegraphics[width=0.4925\columnwidth]{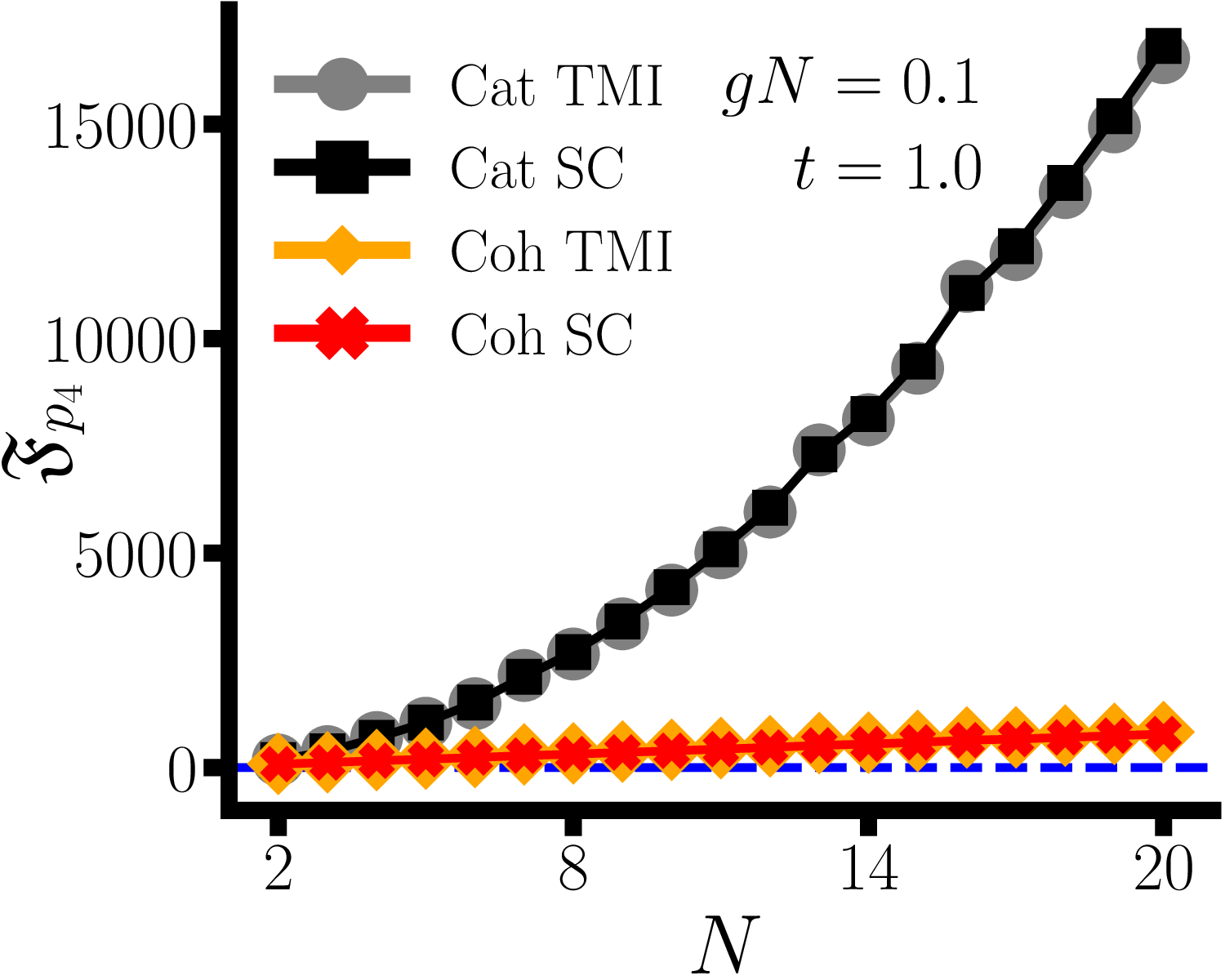}
\includegraphics[width=0.4925\columnwidth]{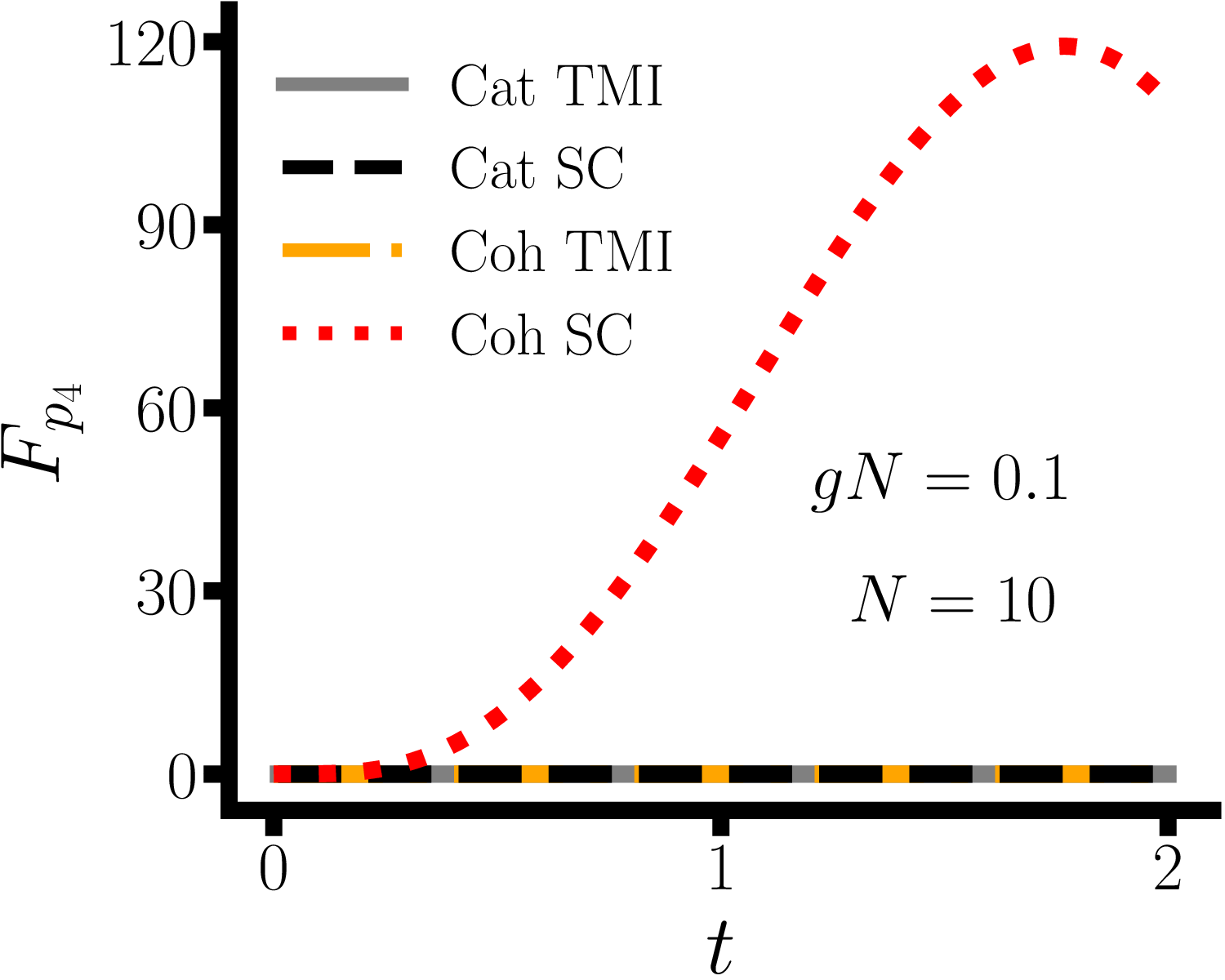}
\includegraphics[width=0.4925\columnwidth]{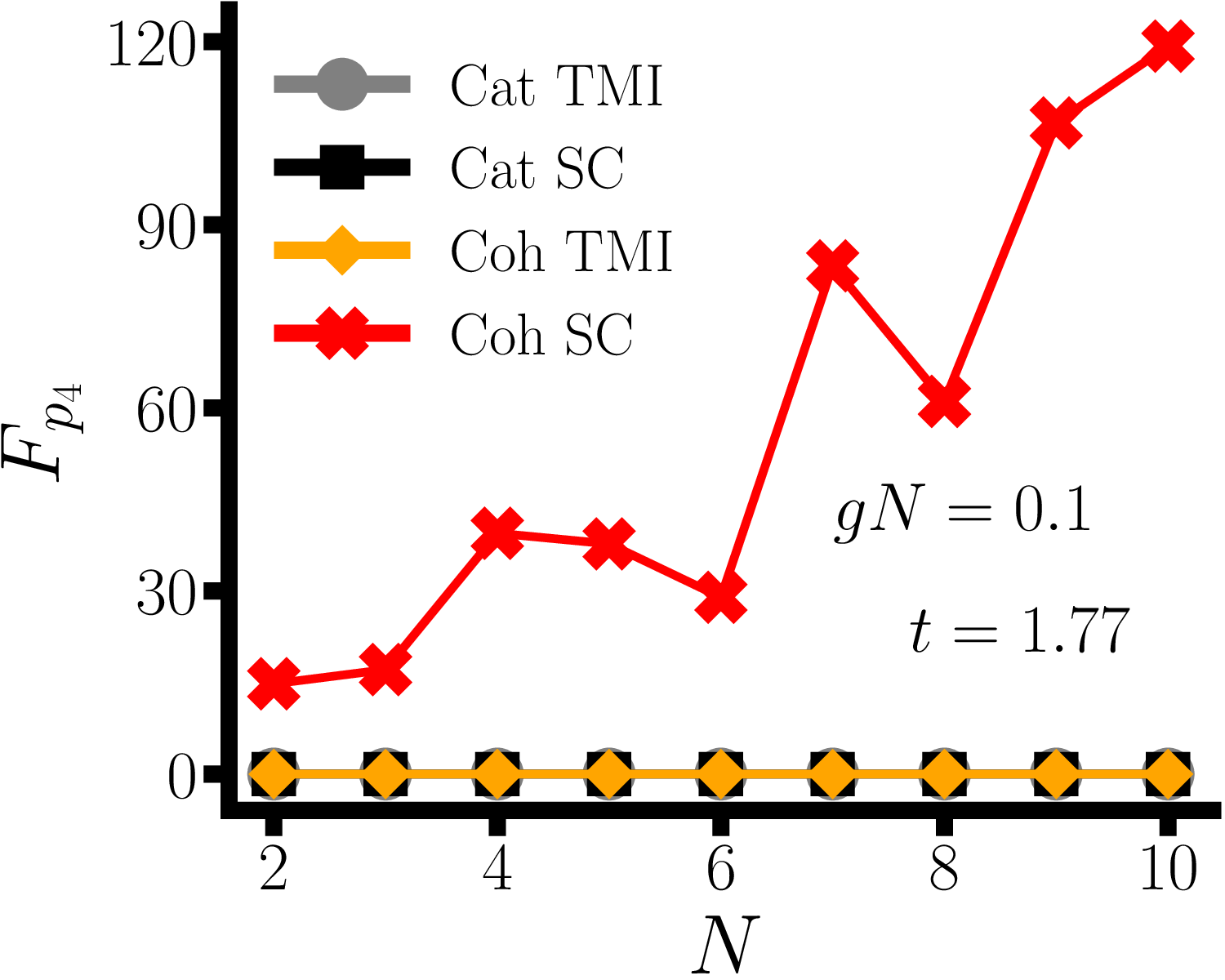}
\caption{\label{fig:qfi}First and second row display quantum ($\mathfrak{F}$) and classical (F) Fisher information, respectively, plotted versus time $t$ (left) and particle number $N$ (right), for cat and coherent (coh) states, respectively. %``TMI'' denotes the conventional two-mode interferometry, and ``SC'' stands for the self-consistent approach.
{At the maximum on the lower left plot, which is at $t=1.77$, the tunneling amplitude in the SC evolution has increased to 
$\tau\simeq 0.09$ (for $N=10$).}}
\end{figure}

We now turn to the CFI \cite{Fisher1922,Fisher1925}, 
associated to a concrete measurement, for which, as we show,   
the impact of self-consistency becomes manifest.
Counting the number of bosons in each well constitutes our measurement,  %l is considered as our measurement outcome,
so that the %corresponding 
CFI is defined as
\begin{equation}
F_{p_4}=\sum_{\vec{\mathbf{n}}}P(\vec{\mathbf{n}}|p_4)\Big(\frac{\partial\log P(\vec{\mathbf{n}}|p_4))}{\partial\,p_4}\Big)^2\,,
\label{eq:cfi}
\end{equation}
where $P(\vec{\mathbf{n}}|p_4)$ is the probability distribution (likelihood) 
 %of the measurement outcomes  
$\vec{\mathbf{n}}=(n_{\rm L},n_{\rm R})$, given $p_4$,
and $n_{\rm L}$ and $n_{\rm R}$ are the numbers of particles that reside in the left and the right well, respectively.
{An appropriate $P(\vec{\mathbf{n}}|p_4)$ has to be constructed to calculate the CFI 
and the conventional TMI approach considers that $P(\vec{\mathbf{n}}|p_4) \coloneqq  
|\langle\vec{n}(t)|\Psi(t)\rangle|^2$.}
Then,  %, within the two-site Bose-Hubbard model, 
the CFI {\em always exactly vanishes,}  
irrespective of the initial state, as the Hamiltonian contains only $\hat{J}_z$ and $\hat{J}_z^2$. 
TMI time evolution therefore just changes the phases of the coefficients: 
$C_k(t)=\exp(-i\epsilon\frac{N-2k}{2}t)\,\exp(-iU(\frac{N-2k}{2})^2t)\,C_k(0)$,
where the state 
$|\psi(t)\rangle=\sum_{k=0}^NC_k(t)|N-k,k\rangle %$,$|\psi(t)\rangle
=\exp(-i(\epsilon\hat{J}_z+U\hat{J}_z^2)t)|\psi(0)\rangle$ 
and $\hat{J}_z|N-k,k\rangle=\frac{N-2k}{2}|N-k,k\rangle$.
Then $P(\vec{\mathbf{n}}=(N-k,k)|p_4)=|C_k(t)|^2=|C_k(0)|^2$, independent of $p_4$, 
yielding vanishing CFI,  %for $p_4$ 
%in the TMI approach, 
{cf.~\cite{suppl}. %, section IV}. %, which explains $0$ CFI.

On the other hand, when both orbitals and %evolve together with the 
Fock space coefficients evolve in the SC framework,
the initial interpretation of the orbitals cannot be maintained throughout the time evolution. 
Initially well-localized orbitals, i.e., $\phi_{\rm L}(x)=\phi_1(x,0)$ and $\phi_{\rm R}(x)=\phi_2(x,0)$, 
%, match enough with the notion of measurement results.
correspond to particles being found in the left and right well,  
%is regarded as belonging to $\phi_{\rm L}(x)$ and $\phi_{\rm R}(x)$, 
respectively.
However, the orbitals change with time during the many-body evolution, so %thus the orbitals at a later time, i.e., 
$\phi_1(x,t)$ and $\phi_2(x,t)$, do not necessarily imply left or right localization at the time of measurement.
Therefore, simply computing 
 $P(\vec{\mathbf{n}}|p_4)=|\langle\vec{n}(t)|\Psi(t)\rangle|^2$, as in the TMI approach, is not applicable. 
In the bosonic field operator $\hat{\Psi}(x)=\sum_j\hat{b}_j(t)\,\phi_j(x,t)$,
it is clear that the bosonic annihilation operator $\hat{b}_j(t)$ corresponds to the time-evolving orbital $\phi_j(x,t)$,
which %may be 
delocalizes with increasing %at some given time 
$t$, see Fig.~\ref{fig:orbitals}.
%In conclusion, 
Thus the Fock state in Eq.~\eqref{eq:ansatz}
%\begin{equation}
$|\vec{n}(t)\rangle=
\frac{\big(\hat{b}_1^\dagger(t)\big)^{n_1}\big(\hat{b}_2^\dagger(t)\big)^{n_2}\cdots\big(\hat{b}_M^\dagger(t)\big)^{n_M}}{\sqrt{n_1!n_2!\cdots n_M!}}
|0\rangle$
%\end{equation}
cannot by itself  
%appropriately 
project the quantum state into any of $|\vec{\mathbf{n}}\rangle$
and $\langle\vec{n}(t)|\Psi(t)\rangle$ cannot be interpreted as probability amplitude for each measurement outcome as in TMI. 
In other words, $|\vec{n}(t)\rangle$ and $|\vec{\mathbf{n}}\rangle$, {while initially identical, become different due to SC evolution. }
Hence we need to re-establish a connection
between time-evolving orbitals and $|\vec{\mathbf{n}}\rangle$ that corresponds to a measurement outcome, resulting
%so that one can obtains 
in the proper distribution $P(\vec{\mathbf{n}}|p_4)=|\langle\vec{\mathbf{n}}|\Psi(t)\rangle|^2 \neq   
%which is not equal to $
|\langle\vec{n}(t)|\Psi(t)\rangle|^2$. 
%This reestablishment will depend on every concrete system.

Within SC time evolution after a trap tilt, %potential is tilted,
$\phi_1(x,t)$ and $\phi_2(x,t)$ remain well-localized in the case of a cat state.
That is, $\phi_1(x,t)$ ($\phi_2(x,t)$) begins from $\phi_1(x,0)=\phi_{\rm L}(x)$ [$\phi_2(x,0)=\phi_{\rm R}(x)$] 
and their absolute value remains nearly identical 
except slightly wider (narrower) width and shorter (taller) height, respectively.
For the spin-coherent state, however, $\phi_1(x,t)$ and $\phi_2(x,t)$ spread  into opposite wells
while they evolve under nonzero $p_4$; see Fig.~\ref{fig:orbitals}.
Then even when a particle resides in $\phi_1(x,t)$ or $\phi_2(x,t)$, to assign it to the 
left or right well is ambiguous.
%In terms of one-body potential with small nonzero $p_4$, however,
One can however still define mathematically ``left’’ or ``right’’
by integrating the orbitals from $-\infty$ to the center point $x=0$ 
of $V(x)$ and from the center point of $V(x)$ to $\infty$, respectively:
$P_\text{Left}=\int_{-\infty}^0\!\!|\phi_j(x,t)|^2\,dx$ and $P_\text{Right}=\int_0^\infty\!\!|\phi_j(x,t)|^2\,dx$. 
%Using these probabilities and bosonic statistics,
One can then construct %the correct 
$P(\vec{\mathbf{n}}|p_4)$ by computing the permanent of a special 
matrix composed from  $P_\text{Left} $ and $P_\text{Right}$. The computable $N$ range 
is limited, though, due to rapidly increasing algorithmic complexity 
{when calculating a matrix permanent, see \cite{suppl}. % section III.} 
%This limitation, however, does not affect the QFI calculation.  %; cf. 
  
% see Appendix \ref{sec:mctdh-probdist} for further details.

The second row in Fig.~\ref{fig:qfi} shows the CFI versus $t$ and $N$. 
%The conventional approach with f
Fixed orbitals results in vanishing CFI, as expected.
Also, even within SC evolution, the cat state shows almost vanishing CFI, 
which is attributed to the fact that the orbitals stay localized in each well during the whole evolution time
and the probabilities, i.e., $P_\text{Left}$ and $P_\text{Right}$, remain nearly constant (for small $p_4$). 
Thus under the given 
measurement the change of $P(\vec{\mathbf{n}}|p_4)$ with respect to $p_4$ is negligible.
However, the spin-coherent state displays a significant change in orbitals
%and displays 
and increasing CFI during the early stage.  % of time evolution. 
Bottom right in Fig.~\ref{fig:qfi} 
shows the $N$-scaling of the CFI; the SC approach with spin-coherent state shows an 
almost linearly increasing CFI.
The complex fluctuation pattern appears because of the short time $t=1.77$ after a nonzero $p_4$ is suddenly applied, and %as the measurement time 
for increasing $t$ % increases %(as well as for increasing values of $gN$) 
these fluctuations smoothen out. %tends to be smoother.
Thus a SC approach may yield drastically different metrological predictions from a TMI based method. 
%and discrepancies from the conventional one in the metrological sense.

\begin{figure}[t]
\includegraphics[width=0.8\columnwidth]{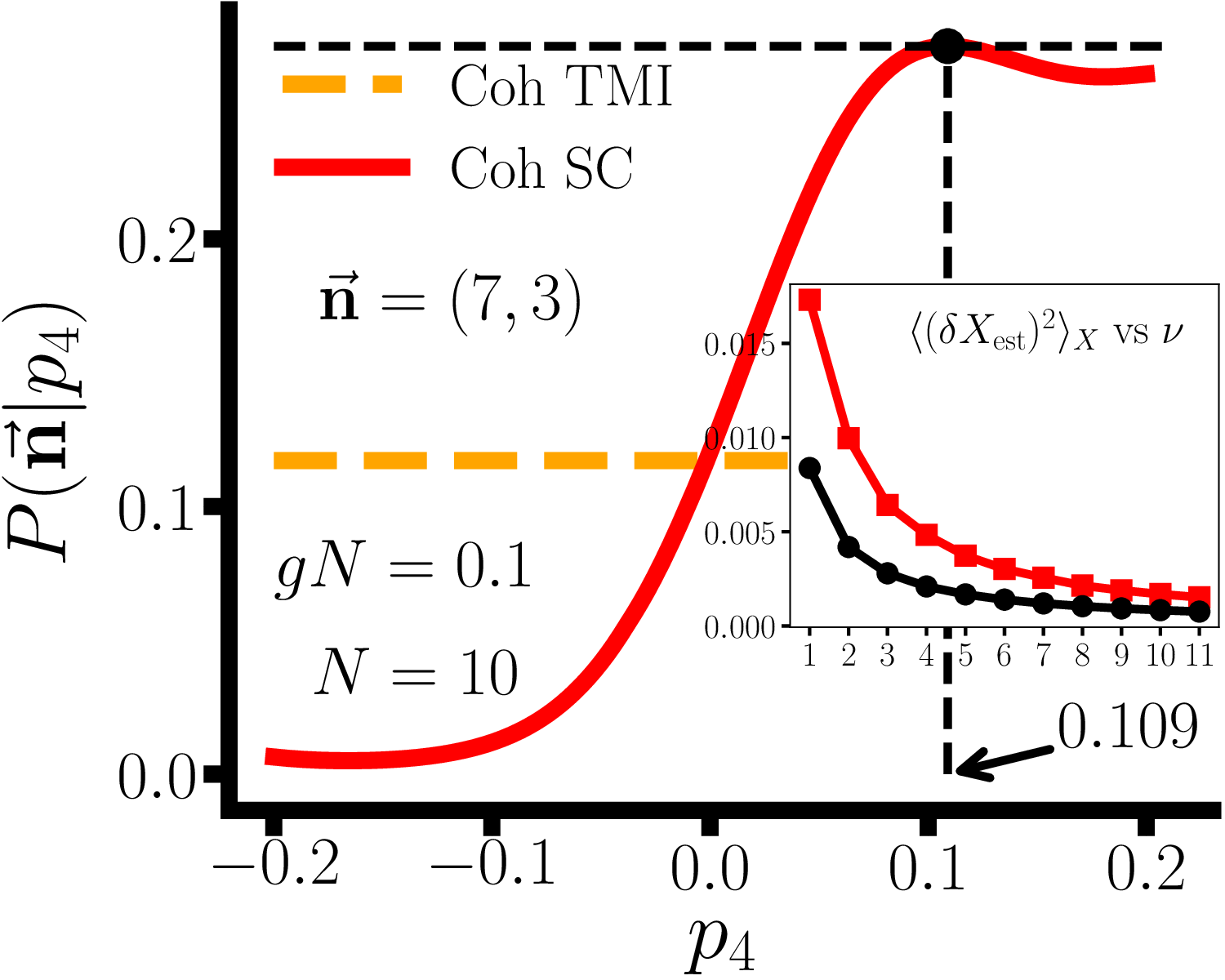}
\caption{\label{fig:mle}Single implementation of the MLE for an estimate of $p_4$ using the 
spin-coherent state. Red solid line is for the  SC approach and orange dashed line (a constant $\forall \,p_4$) for the TMI.
In the inset, red squares represent the mean-square deviation and black rounds the Cram\'er-Rao lower bound
from Eq.~\eqref{eq:cramér-rao}.}
\end{figure}

Fig.~\ref{fig:mle} shows our primary result: The implementation of parameter 
estimation at the final stage of the metrological protocol.
%Since the spin-coherent state shows nonzero CFI in the SC framework,
%we use only that state here.
The maximum likelihood estimator (MLE) \cite{Howard2010,Rossi} 
is used as a concrete example,  
{because it asymptotically saturates the Cramér-Rao bound 
for an infinite number of measurements  ($\nu\rightarrow\infty$, see also
%to obtain a single estimate 
\cite{suppl}) %, section V), 
%\cite{Braunstein_1994} %,Howard2010}:
\begin{equation}
\langle(\delta X_{\text{est}})^2\rangle_X\geq\frac{1}{\nu F_X}\,.
\label{eq:cramér-rao}
\end{equation}
Here, the parameter $X$ is our $p_4$, $X_{\text{est}}$ is an estimator of $X$,
and %its deviation is 
$\delta X_{\text{est}}:=X_{\text{est}}/\big|\partial\langle X_{\text{est}}\rangle_X/\partial X\big|-X$, with 
average $\langle\cdots \rangle_X $ taken with respect to $P(\vec{\mathbf{n}}|p_4)$.
%The asymptotic saturation of Eq.~\eqref{eq:cramér-rao} by the MLE implies that equality in the bound 
%increasingly 
%is attained when $\nu\rightarrow\infty$ \cite{suppl}. 

In Fig.~\ref{fig:mle}, the likelihood function $P(\vec{\mathbf{n}}|p_4)$ %versus $p_4$ 
is displayed, supposing, for concreteness, that the measurement outcome is $\vec{\mathbf{n}}
%=(n_{\rm L},n_{\rm R})
=(7,3)$, where
$n_{\rm L}$ and $n_{\rm R}$ denote number of particles %found 
in left and right well, respectively.
%,that is, 7 particles are found in the left well and 3 found in the right well.
Red solid line shows the maximum of $P(\vec{\mathbf{n}}|p_4)$ at $p_4\simeq0.109$,
thus the estimate of $p_4$, given $\vec{\mathbf{n}}=(7,3)$, is $X_{\text{est}}\simeq0.109$.
Similarly, every single outcome, $11$ in total, is connected to an % corresponding 
estimate of $p_4$.
The orange dashed line obtained by %the %that corresponds to the 
conventional TMI %approach 
with a spin-coherent state
stays flat, which means that the information given by the measurement outcome $\vec{\mathbf{n}}=(7,3)$ is zero, 
%so one cannot extract 
and an estimate of $p_4$, in accordance with $F_{p_4}=0$, is not possible. 

When $\langle X_{\text{est}}\rangle_X=X$ holds, an estimator is %by definition 
unbiased. 
The present MLE is not unbiased except infinitesimally close to $p_4=0$.  %, and it depends on each circumstance.
%Unbiasedness
%This is  assessed by $\langle X_{\text{est}}\rangle_X$}
%In our current example, this is approximately fullfilled near $p_4=0$,
The larger  $p_4$, the more bias its estimation acquires. For instance, in Fig.~\ref{fig:mle}, 
the true value of $p_4$ is $0.1$ and the bias of the MLE is 
on the level of 10\,\%, 
%slightly biased, % such that 
$\langle X_{\text{est}}\rangle_{X=0.1}\simeq0.11$. 
{To compensate for this bias, we calibrate the MLE by obtaining $\langle X_{\text{est}}\rangle (X)$ from our MCTDH simulations, see \cite{suppl}.} %, section V.}
Also, $\partial\langle X_{\text{est}}\rangle_X/\partial X|_{X=0.1}\simeq0.6$, 
%Then all values are ready to calculate the 
which then provides %enables to calculate 
the mean-square deviation $\langle(\delta X_{\text{est}})^2\rangle_{X=0.1}$. 
The inset in Fig.~\ref{fig:mle} verifies that, asymptotically, %as it should,   
%the mean-square deviation 
$\langle(\delta X_{\text{est}})^2\rangle_{X=0.1}$ 
approaches the Cramér-Rao lower bound % (CRLB),
%i.e., the right hand side in Eq.~\eqref{eq:cramér-rao}, 
for  %increasing 
large $\nu$.  %of measurement outcomes 
%\col{which appears in Eq.~\eqref{eq:cramér-rao} due to the central limit theorem.} 
%; we refer to \cite{suppl} 
%Refer to Appendix.~\ref{sec:mctdh-mle} 
%for details on the MLE.

In conclusion, we have found, using a self-consistent many-body approach,  
% As long as the MCTDH theory is regarded as more realistic simulation of the ultracold atomic gas
%under limited computational resources or approximate theory,
that many-body metrology %the % interpretation of the 
%metrological outcome of a given measurement 
utilizing interacting trapped bosons 
%interpretation %of any given input state 
%at the instant of measurement 
needs to conform to the final self-consistently computed many-body state. 
% that is based on such simulated quantum state to the MCTDH framework.
%On one hand, t
The QFI completely depends on the parameter dependence of the final state itself,
and is thus relatively unaffected by self-consistency (in the weakly interacting regime).
%This justifies the use of conventional TMA with the time-independent Fock basis.
However, even in the latter regime, the CFI for a  %a concrete 
parameter estimation experiment %, as we have shown, 
is strongly affected by self-consistency due to its sensitive dependence 
%because the measurement results that experimenters attain may rely on 
on the orbitals' time evolution. 
As a particularly notable example, 
%we have shown that 
fitting the outcome of a number-statistics experiment in a double well 
% measuring e.g. a density profile, 
to conventional TMI %would yield %s,  %(having vanishing CFI) yields, 
gives a null result for estimating the slope parameter $p_4$. 
%would allow for no parameter estimation,  
%while fitting the results of 
The SC approach we employ, however, 
%, also kept within the confines of the TMA, 
enables $p_4$ estimation. 
%As a corollary, the very accuracy of the MCTDH
%approach delivering the predicted density profile can be assessed}
Metrology with trapped ultracold quantum gases thus in general requires 
self-consistency of dynamical evolution, to correctly predict the estimation precision that can be 
accomplished in a given metrological %matter wave 
protocol. 

%\begin{acknowledgments}
This work has been supported by the National Research Foundation of Korea under 
Grants No.~2017R1A2A2A05001422 and No.~2020R1A2C2008103.
%\end{acknowledgments}

% Specify following sections are appendices. Use \appendix* if there
% only one appendix.
%\appendix
\bibliography{metro9}

\begin{thebibliography}{58}
\bibitem [{exp()}]{explanation}%
  \BibitemOpen
  \href@noop {} {}\bibinfo {note} {This may be explained by the fact that terms
  containing Fock space coefficients only and those involving orbitals in
  ${\mathfrak F}_X$ (Eq.~(S4) in \cite{suppl}) tend to compensate each
  other.}\BibitemShut {Stop}%
\end{thebibliography}

%\end{document}

%\vspace*{10em} 
%\newpage 
%\appendix
\newpage
%\pagebreak 

%\begin{widetext}

\begin{widetext}
\setcounter{equation}{0}
\setcounter{figure}{0}
\setcounter{table}{0}
\setcounter{page}{1}
\renewcommand{\theequation}{S\arabic{equation}}
\renewcommand{\thefigure}{S\arabic{figure}}

\section{Supplemental Material}

\subsection{\label{sec:mctdh-theory}I. Multiconfigurational time-dependent Hartree theory}

Given a set of coefficients and a set of orbitals for an initial state,
the time evolution in the MCTDH framework proceeds according to the following system of equations:
\begin{eqnarray}
i\,\frac{\partial\,\mathbf{C}(t)}{\partial\,t}&=&\mathbf{H}(t)\,\mathbf{C}(t)\,,\nonumber\\
i\,\partial_t|\phi_j\rangle&=&\hat{P}\Big[\hat{h}|\phi_j\rangle+\sum_{k,s,q,l}[\rho^{-1}]_{jk}\rho_{ksql}\hat{W}_{sl}|\phi_q\rangle\Big],
\label{eq:evolution}
\end{eqnarray}
which is derived by applying the time-dependent variational principle to the interacting $N$-body Hamiltonian
$\hat{H}=\sum_{j=1}^N\hat{h}(x_j)+\sum_{j<k}\hat{W}(x_j-x_k)$\cite{Ofir2008}.
Here, $\mathbf{C}(t)$ is a column vector that consists of all possible expansion coefficients $C_{\vec{n}}(t)$
and $\mathbf{H}(t)$ corresponds to the time-dependent Hamiltonian matrix in the basis $\{|\vec{n}(t)\rangle\}$.
Also, $\hat{h}$ is a single-particle Hamiltonian,
$\hat{W}_{sl}=\int dx'\,\phi_s^\ast(x')\,\hat{W}(x-x')\,\phi_l(x')$,
and $\hat{P}=1-\sum_{j=1}^M|\phi_j\rangle\langle\phi_j|$ is an projection operator to the subspace
that is orthogonal to the one spanned by orbitals.
The $[\rho^{-1}]_{jk}$ is a matrix element of the inverse of reduced one-body density matrix:
\begin{eqnarray}
\rho(x,x';t)&=&\langle\Psi(t)|\hat{\Psi}^\dagger(x')\hat{\Psi}(x)|\Psi(t)\rangle\nonumber\\
&=&\sum_{k,q}\phi_k^\ast(x',t)\phi_q(x,t)\langle\Psi(t)|\hat{b}_k^\dagger(t)\hat{b}_q(t)|\Psi(t)\rangle\nonumber\\
&=&\sum_{k,q}\phi_k^\ast(x',t)\phi_q(x,t)\rho_{kq}(t)\,,
\label{eq:robdm}
\end{eqnarray}
where the $\rho_{kq}$ is, for the cases of $k=q$ and $k\neq q$,
\begin{equation*}
\rho_{kk}=\sum_{\vec{n}}|C_{\vec{n}}(t)|^2n_k,\quad
\rho_{kq}=\sum_{\vec{n}}C_{\vec{n}}^\ast(t)C_{\vec{n}_k^q}(t)\sqrt{n_k(n_q+1)}\,.
\end{equation*}
Similarly, $\rho_{ksql}$ is a matrix element of the reduced two-body matrix
\begin{eqnarray}
\rho(x_1,x_2,x'_1,x'_2;t)&=&\langle\Psi(t)|\hat{\Psi}^\dagger(x'_1)\hat{\Psi}^\dagger(x'_2)\hat{\Psi}(x_1)\hat{\Psi}(x_2)|\Psi(t)\rangle\nonumber\\
&=&\sum_{k,s,q,l}\phi_k^\ast(x'_1,t)\phi_s^\ast(x'_2,t)\phi_q(x_1,t)\phi_l(x_2,t)\rho_{ksql}(t)\,,
\label{eq:rtbdm}
\end{eqnarray}
where
\begin{eqnarray*}
\rho_{kkkk}&=&\sum_{\vec{n}}|C_{\vec{n}}(t)|^2n_k(n_k-1)\,,\quad
\rho_{ksks}=\sum_{\vec{n}}|C_{\vec{n}}(t)|^2n_kn_s\,,\nonumber\\
\rho_{kkqq}&=&\sum_{\vec{n}}C_{\vec{n}}^\ast(t)\,C_{\vec{n}_{kk}^{qq}}(t)\sqrt{(n_k-1)n_k(n_q+1)(n_q+2)}\,,\quad
\rho_{kkkl}=\sum_{\vec{n}}C_{\vec{n}}^\ast(t)\,C_{\vec{n}_k^l}(t)(n_k-1)\sqrt{n_k(n_l+1)}\,,\nonumber\\
\rho_{ksss}&=&\sum_{\vec{n}}C_{\vec{n}}^\ast(t)\,C_{\vec{n}_k^s}(t)n_s\sqrt{n_k(n_s+1)}\,,\quad
\rho_{kkql}=\sum_{\vec{n}}C_{\vec{n}}^\ast(t)\,C_{\vec{n}_{kk}^{ql}}(t)\sqrt{(n_k-1)n_k(n_q+1)(n_l+1)}\,,\nonumber\\
\rho_{ksqq}&=&\sum_{\vec{n}}C_{\vec{n}}^\ast(t)\,C_{\vec{n}_{ks}^{qq}}(t)\sqrt{n_kn_s(n_q+1)(n_q+2)}\,,\quad
\rho_{kssl}=\sum_{\vec{n}}C_{\vec{n}}^\ast(t)\,C_{\vec{n}_k^l}(t)n_s\sqrt{n_k(n_l+1)}\,,\nonumber\\
\rho_{ksql}&=&\sum_{\vec{n}}C_{\vec{n}}^\ast(t)\,C_{\vec{n}_{ks}^{ql}}(t)\sqrt{n_kn_s(n_q+1)(n_l+1)}\,.
\end{eqnarray*}

Infinite resources for numerical calculation makes it possible to assume the theoretical limit $M\rightarrow\infty$,
thus $\hat{P}\rightarrow\hat{0}$ and $\partial_t|\phi_j\rangle=0$ in Eq.~\eqref{eq:evolution},
which means that a complete set of time-independent orbitals $\{\phi_j(x)\,|\,j=1,2,\cdots\}$ can be composed
and the dynamics of systems is fully described only by the set of coefficients $\{C_{\vec{n}}(t)\}$,
from which all quantum metrological properties can be extracted.

If $M=2$, with fixed orbitals,  is adequate for the description of a system, the modes comprise the conventional TMI,  %two-mode interferometer 
using a  SU(2) formulation.
Optical systems have been used to realize such two-mode systems, e.g., a Mach-Zehnder interferometer,
where only the Fock space coefficients matter to predict the number of photons in each interferometric arm.
However, for  interacting atoms, an exact description requires infinite $M$,
and truncating at finite $M$ is valid only approximately, cf. the error-controlled extension of 
multiconfigurational Hartree put forth in~\cite{Kang-Soo}. 
As the interaction becomes weaker, a description in terms of finite $M$ improves. 
The self-consistent MCTDH framework here goes significantly further 
further than a conventional TMI and introduces time-evolving orbitals of changing shape.{We also note here %, regarding the latter fact, 
that a Hartree-Fock method, using plane waves 
for the field operator expansion as appropriate in a translationally invariant system, will fail to capture
a trapped system when, as necessary for finite computational resources, the %field operator 
expansion is  truncated at a finite $M$.}

%so that the limited number of modes can simulate the cold atomic or molecular systems with improved exactness.

\subsection{\label{sec:mctdh-qfi}II. Quantum Fisher information of a pure state in the MCTDH framework}

Because of the introduction of time-evolving orbitals,
a formulation of the QFI is required which facilitates incorporating the result %physical quantities
%calculated in the process 
of solving the MCTDH time evolution in Eqs.~\eqref{eq:evolution}. 
The QFI, which is the ultimate limit of precision given by $|\psi_X\rangle$, is calculated by
$\mathfrak{F}_X=4\big(\langle\partial_X\psi_X|\partial_X\psi_X\rangle
-|\langle\psi_X|\partial_X\psi_X\rangle|^2\big)$ for  general pure states,
and for some state represented as Eq.~\eqref{eq:ansatz}, we have, for any number of modes, 
\begin{eqnarray}
\mathfrak{F}_X/4&=&\sum_{\vec{n}}\partial_X C_{\vec{n}}^\ast\,\partial_X C_{\vec{n}}
-\Big|\sum_{\vec{n}}C_{\vec{n}}^\ast\,\partial_X C_{\vec{n}}\Big|^2\nonumber\\
&&\,\,+\sum_{\vec{n}}\sum_{k,q}
\big(\partial_X C_{\vec{n}}^\ast\,C_{\vec{n}_k^q}-C_{\vec{n}}^\ast\,\partial_X C_{\vec{n}_k^q}\big)
(\partial_X)_{kq}\,\zeta_{qk}-\sum_{\vec{n}}\big(\partial_X C_{\vec{n}}^\ast\,C_{\vec{n}}
-C_{\vec{n}}^\ast\,\partial_X C_{\vec{n}}\big)\sum_{k,q}(\partial_X)_{kq}\,\rho_{kq}\nonumber\\
&&\quad-\sum_{k,s,q}(\partial_X)_{ks}(\partial_X)_{sq}\,\rho_{kq}+\Big(\sum_{k,q}(\partial_X)_{kq}\rho_{kq}\Big)^2
-\!\sum_{k,s,q,l}(\partial_X)_{kq}(\partial_X)_{sl}\,\rho_{ksql}\,,
\label{eq:qfi}
\end{eqnarray}
where $\zeta_{qk}:=\sqrt{n_k(n_q+1)}$ or $\zeta_{qk}:=n_k$ if $q\neq k$ or $q=k$, respectively,
and $(\partial_X)_{kq}:=\int dx\,\phi^\ast_k(x,t)\,\partial_X\phi_q(x,t)$.
Refer to Eq.~\eqref{eq:robdm} and Eq.~\eqref{eq:rtbdm} for the definitions of $\rho_{kq}$ and $\rho_{ksql}$.
The first two terms involve only the coefficients and the remaining terms are related to the changes of coefficients and orbitals, 
for infinitesimal increment of $X$. In summary, Eq.~\eqref{eq:qfi}  completely 
incorporates the information orbitals as well as coefficients changing with $X$. 
%, beyond the first line which represents the contribution of conventional TMI.}

\subsection{\label{sec:mctdh-probdist}III. Construction of the probability distribution (likelihood) 
in MCTDH for bosons}

Here we explain how to construct the probability distribution of measurement outcomes.
This process obviously depends on the specific systems and the choice of measurement.
Here, the metrological implementation with the ultracold bosons trapped in a double-well potential is covered
and the number of particles in each well is counted after the time evolution is finished, and considered as the measurement.
%At the time of measurement, the barrier in the double-well potential isolates enough the left and the right,
%thus t
The probability of a particle in $\phi_j(x,t)$ to be found at the left ($L$) or the right ($R$) is defined as 
\begin{equation}
P_{L,j}=\int_{-\infty}^0\!\!|\phi_j(x,t)|^2\,dx\,,\qquad
P_{R,j}=\int_0^{\infty}\!\!|\phi_j(x,t)|^2\,dx\,,
\end{equation}
where we assume that the center of the 1D potential is at $x=0$.

Next, we need to consider the combinatorial problem related to many particles and bosonic statistics.
Let us take for simplicity the example of $N=2$.
There are three measurement outcomes: $\vec{\mathbf{n}}:=(n_{\rm L},n_{\rm R})=(2,0)$, $(1,1)$, and $(0,2)$,
in which $n_{\rm L}$ and $n_{\rm R}$ mean the numbers of particles found in the left well and in the right well, respectively.
When the final state is $\sum_{\vec{n}}C_{\vec{n}}|\vec{n}\rangle=\sum_{k=0}^2C_k|2-k,k\rangle$, the probability for each case is as follows:
\begin{eqnarray}
P_0&=&P\big(\vec{\mathbf{n}}=(2,0)\big)=|C_0|^2P_{{\rm L},1}^2+|C_1|^2P_{{\rm L},1}P_{{\rm L},2}+|C_2|^2P_{{\rm L},2}^2\,,\nonumber\\
P_1&=&P\big(\vec{\mathbf{n}}=(1,1)\big)=2\,|C_0|^2P_{{\rm L},1}P_{{\rm R},1}+|C_1|^2(P_{{\rm L},1}P_{{\rm R},2}+P_{{\rm R},1}P_{{\rm L},2})+2\,|C_2|^2P_{{\rm L},2}P_{{\rm R},2}\,,\nonumber\\
P_2&=&P\big(\vec{\mathbf{n}}=(0,2)\big)=|C_0|^2P_{{\rm R},1}^2+|C_1|^2P_{{\rm R},1}P_{{\rm R},2}+|C_2|^2P_{{\rm R},2}^2\,,
\end{eqnarray}
where it is trivial to show that $P(\vec{\mathbf{n}}=(2,0))+P(\vec{\mathbf{n}}=(1,1))+P(\vec{\mathbf{n}}=(0,2))=1$
using $P_{j,L}+P_{j,R}=1$ and $|C_0|^2+|C_1|^2+|C_2|^2=1$.
After careful inspection, one can rewrite the above probabilities as
\begin{eqnarray}
P_0&=&P\big(\vec{\mathbf{n}}=(2,0)\big)=\frac{|C_0|^2}{2}
\Big\{\begin{array}{cc}
P_{{\rm L},1} & P_{{\rm L},1} \\
P_{{\rm L},1} & P_{{\rm L},1}
\end{array}\Big\}
+\frac{|C_1|^2}{2}
\Big\{\begin{array}{cc}
P_{{\rm L},1} & P_{{\rm L},2} \\
P_{{\rm L},1} & P_{{\rm L},2}
\end{array}\Big\}
+\frac{|C_2|^2}{2}
\Big\{\begin{array}{cc}
P_{{\rm L},2} & P_{{\rm L},2} \\
P_{{\rm L},2} & P_{{\rm L},2}
\end{array}\Big\},\nonumber\\
P_1&=&P\big(\vec{\mathbf{n}}=(1,1)\big)=|C_0|^2
\Big\{\begin{array}{cc}
P_{{\rm L},1} & P_{{\rm L},1} \\
P_{{\rm R},1} & P_{{\rm R},1}
\end{array}\Big\}
+|C_1|^2
\Big\{\begin{array}{cc}
P_{{\rm L},1} & P_{{\rm L},2} \\
P_{{\rm R},1} & P_{{\rm R},2}
\end{array}\Big\}
+|C_2|^2
\Big\{\begin{array}{cc}
P_{{\rm L},2} & P_{{\rm L},2} \\
P_{{\rm R},2} & P_{{\rm R},2}
\end{array}\Big\},\nonumber\\
P_2&=&P\big(\vec{\mathbf{n}}=(0,2)\big)=\frac{|C_0|^2}{2}
\Big\{\begin{array}{cc}
P_{{\rm R},1} & P_{{\rm R},1} \\
P_{{\rm R},1} & P_{{\rm R},1}
\end{array}\Big\}
+\frac{|C_1|^2}{2}
\Big\{\begin{array}{cc}
P_{{\rm R},1} & P_{{\rm R},2} \\
P_{{\rm R},1} & P_{{\rm R},2}
\end{array}\Big\}
+\frac{|C_2|^2}{2}
\Big\{\begin{array}{cc}
P_{{\rm R},2} & P_{{\rm R},2} \\
P_{{\rm R},2} & P_{{\rm R},2}
\end{array}\Big\},
\end{eqnarray}
in which $\{V\}$ means the permanent of a matrix $V$.
By tracing the factor in front of each term and by considering bosonic statistics,
one can find a regular pattern and generalize as follows:
\begin{equation}
P_j=P\big(\vec{\mathbf{n}}=(N-j,j)\big)=\frac{1}{N!}
\Big(\begin{array}{c}
N \\
j
\end{array}\Big)
\sum_{k=0}^{N}|C_k|^2
\{V_{j,k}\}, \label{eq:scprob}
\end{equation}
where $V_{j,k}$ is a special $N\times N$ matrix, defined as below. 
The $j$ is the number of particles in the right well,
i.e, $\vec{\mathbf{n}}=(N-j,j)$ and the $k$ means $\vec{n}=(N-k,k)$.
The example above shows how to  compose the matrix $V_{j,k}$.
In order to compose $V_{1,2}$, for example, ``$1$" is represented as $\{L, R\}$
and the ``$2$" is represented as $\{2,2\}$.
The former is an ordered set of $N-j$ of $L$ and $j$ of $R$,
and the latter is a conversion of ``how many particles there are in each mode''
into an (ascending-)ordered set of the occupied mode numbers:
\begin{eqnarray*}
j=0\,:\,(2,0)\quad\rightarrow\quad\{L,L\}\,,\qquad
j=1\,:\,(1,1)\quad\rightarrow\quad\{L,R\}\,,\qquad
j=2\,:\,(0,2)\quad\rightarrow\quad\{R,R\}\,,\\~\\
k=0\,:\,(2,0)\quad\rightarrow\quad\{1,1\}\,,\qquad
k=1\,:\,(1,1)\quad\rightarrow\quad\{1,2\}\,,\qquad
k=2\,:\,(0,2)\quad\rightarrow\quad\{2,2\}\,.
\end{eqnarray*}
Then the former set makes up the row indices and the latter set makes up the column indices:
\begin{eqnarray}
\begin{array}{ccc}
  & 2 & 2 \\
L &   &   \\
R &   &
\end{array}
\quad\rightarrow\quad
\left(\begin{array}{cc}
P_{{\rm L},2} & P_{{\rm L},2} \\
P_{{\rm R},2} & P_{{\rm R},2}
\end{array}\right)
=V_{1,2}\,,
\end{eqnarray}
and its permanent is now readily obtained to be 
\begin{eqnarray}
\{V_{1,2}\}=\Big\{\begin{array}{cc}
P_{{\rm L},2} & P_{{\rm L},2} \\
P_{{\rm R},2} & P_{{\rm R},2}
\end{array}\Big\}
=P_{{\rm L},2}\,P_{{\rm R},2}+P_{{\rm L},2}\,P_{{\rm R},2}\,.
\end{eqnarray}

For another example, let us suppose that $N=3$ and try to express $V_{1,2}$.
The first subscript $1$ is converted into $\{L,L,R\}$
and the second one $2$ is converted into $\{1,2,2\}$.
Then
\begin{eqnarray}
\begin{array}{cccc}
  & 1 & 2 & 2 \\
L &   &   &   \\
L &   &   &   \\
R &   &   &
\end{array}
\quad\rightarrow\quad
\left(\begin{array}{ccc}
P_{{\rm L},1} & P_{{\rm L},2} & P_{{\rm L},2} \\
P_{{\rm L},1} & P_{{\rm L},2} & P_{{\rm L},2} \\
P_{{\rm R},1} & P_{{\rm R},2} & P_{{\rm R},2}
\end{array}\right)
=V_{1,2}\,,
\end{eqnarray}
and the permanent is $\{V_{1,2}\}=4P_{{\rm L},1}P_{{\rm L},2}P_{{\rm R},2}+2P_{{\rm L},2}^2P_{{\rm R},1}$. 
Now we have all ingredients to construct the probability distribution of a measurement 
for which the number of particles in each well is counted. {To calculate the permanent
of a matrix, we used the advanced algorithm developed to reduce the algorithmic complexity 
in \cite{NijenhuisWilf2014}; for an introduction see \cite{Xuewei2020}.}

\subsection{\label{sec:mctdh-mle}IV. Additional details on the %implementation of the 
MLE}
\subsubsection{Construction of estimator and likelihood function} 
The maximum likelihood estimator (abbreviated already in the main text as MLE) 
is a commonly used estimator in the field of statistics and is defined as follows:
\begin{equation}
X_{\text{est}}=\text{argmax}_XP(\vec{\mathbf{n}}|X)\,,
\label{eq:mle}
\end{equation}
where $\vec{\mathbf{n}}$ is used to denote the  measurement outcome. Also, 
argmax$_X$ denotes, by definition of the MLE, the {unique} point %or points, or elements, 
in the domain of interest, %  function 
at which the function values are maximized. 
Whenever an outcome $\vec{\mathbf{n}}$ is attained, one inserts it into the RHS of Eq.~\eqref{eq:mle}
and finds a value of $X$ that maximizes $P(\vec{\mathbf{n}}|X)$.
This is  a one-shot estimate of $X$, namely $X_\text{est}$.
In order to implement the MLE, it is necessary to obtain the likelihood function, i.e., $P(\vec{\mathbf{n}}|X)$, {
the process of which we now describe.}

In the conventional TMI, only considering the change of Fock space coefficients,
$P(\vec{\mathbf{n}}|X)$ is calculated as $P(\vec{n}|X)=\langle\vec{n}|\hat{\Psi}(t)\rangle=|C_{\vec{n}}(t)|^2$,
where $|\Psi(t)\rangle=\sum_{\vec{n}}C_{\vec{n}}(t)|\vec{n}\rangle$. % (recall that $X=p_4$).
For the two-mode (double-well) system covered in the main text, cf.~Eq.\eqref{eq:bose-hubbard},
we may consider the general two-mode state $|\Psi(t)\rangle=\sum_{k=0}^NC_k(t)|N-k,k\rangle$.
Then $\vec{n}=(n_1,n_2)$, denoting  that $n_1$ particles are in $\phi_1(x,t)$ and $n_2$ particles in $\phi_2(x,t)$,
is identified as the measurement result that $n_1$ particles are in the left well and $n_2$ particles are in the right well:
$\vec{\mathbf{n}}=(n_{\rm L}=n_1,n_{\rm R}=n_2)$.
In particular, with the metrological protocol adopted in the main text, i.e., $|\Psi(t)\rangle=e^{-i(\epsilon\hat{J}_z+U\hat{J}_z^2)t}|\Psi(0)\rangle$,
each $C_k(t)$ changes only by a phase (but not by magnitude): 
$C_k(t)=\exp(-i\epsilon\frac{N-2k}{2}t)\,\exp(-iU(\frac{N-2k}{2})^2t)\,C_k(0)$,
where $\epsilon$ contains the information of $X$. 
Hence the likelihood $P(\vec{\mathbf{n}}|X)$ is independent of $X$ and invariant,
which leads to vanishing CFI, see also the constant orange dashed line (coh TMI) in Fig.~\ref{fig:mle}.

In the self-consistent approach, % with finite number of modes, 
however,
the calculation of $P(\vec{\mathbf{n}}|X)$ depends on the specifics of 
each system and measurement  considered, since the measurement results are affected by the changing orbitals as well as by the changing Fock space coefficients.
A quantum state is now written as $|\Psi(t)\rangle=\sum_{\vec{n}}C_{\vec{n}}(t)|\vec{n}(t)\rangle$,
indicating that  the orbitals associated by the Fock space basis state $|\vec{n}(t)\rangle$ evolve in time.
In our double-well system, now $\vec{n}=(n_1,n_2)$ cannot be interpreted as
``$n_1$ particles in the left well and $n_2$ particles in the right well'' anymore.  
The correct statement now is ``$n_1$ particles are in $\phi_1(x,t)$ and $n_2$ particles are in $\phi_1(x,t)$''.
The orbitals may delocalize as time passes, thus at the instant of measurement
a particle in the orbital $\phi_1(x,t)$ can be found
in the left well or in the right well with some probabilities $P_{{\rm L},1}$ or $P_{{\rm R},1}$, respectively,
where $P_{L,j}:=\int_{-\infty}^0|\phi_j(x,t)|^2dx$ and $P_{R,j}:=\int_0^\infty|\phi_j(x,t)|^2dx$.
In other words, at time $t$, it is necessary to take further probability distributions 
 into account other than just $|C_k(t)|^2$:
\begin{eqnarray}
P(\vec{\mathbf{n}}=(N-j,j)|X)&=&|C_k(t)|^2\,,\qquad\qquad\qquad\,\,\,\,\text{Conventional Two-Mode Interferometry} \label{NSC_P}\\
P(\vec{\mathbf{n}}=(N-j,j)|X)&=&\sum_{k=0}P_{j,k}|C_k(t)|^2\,,\qquad\qquad\qquad\quad\text{Self-Consistent Approach} \label{SC_P}
\end{eqnarray}
where the probability coefficients in \eqref{SC_P} read 
\[
P_{j,k}\coloneqq\frac{1}{N!}
\left(\begin{array}{c}
N \\
j
\end{array}\right)
\{V_{j,k}\}\,.
\]
We refer to Eq.~\eqref{eq:scprob} and the discussion it follows 
%\col{there is no such equation in the suppl}
%in the supplement 
for the definition and calculation of the special matrix $\{V_{j,k}\}$.
In summary, the difference in obtaining the probabilities $P(\vec{\mathbf{n}}|X)$ as outlined in the above  
leads to a discrepancy in the probability distribution (synonymously likelihood), and therefore in 
the CFI and the  MLE.

\begin{figure}[b]
\includegraphics[width=0.4\columnwidth]{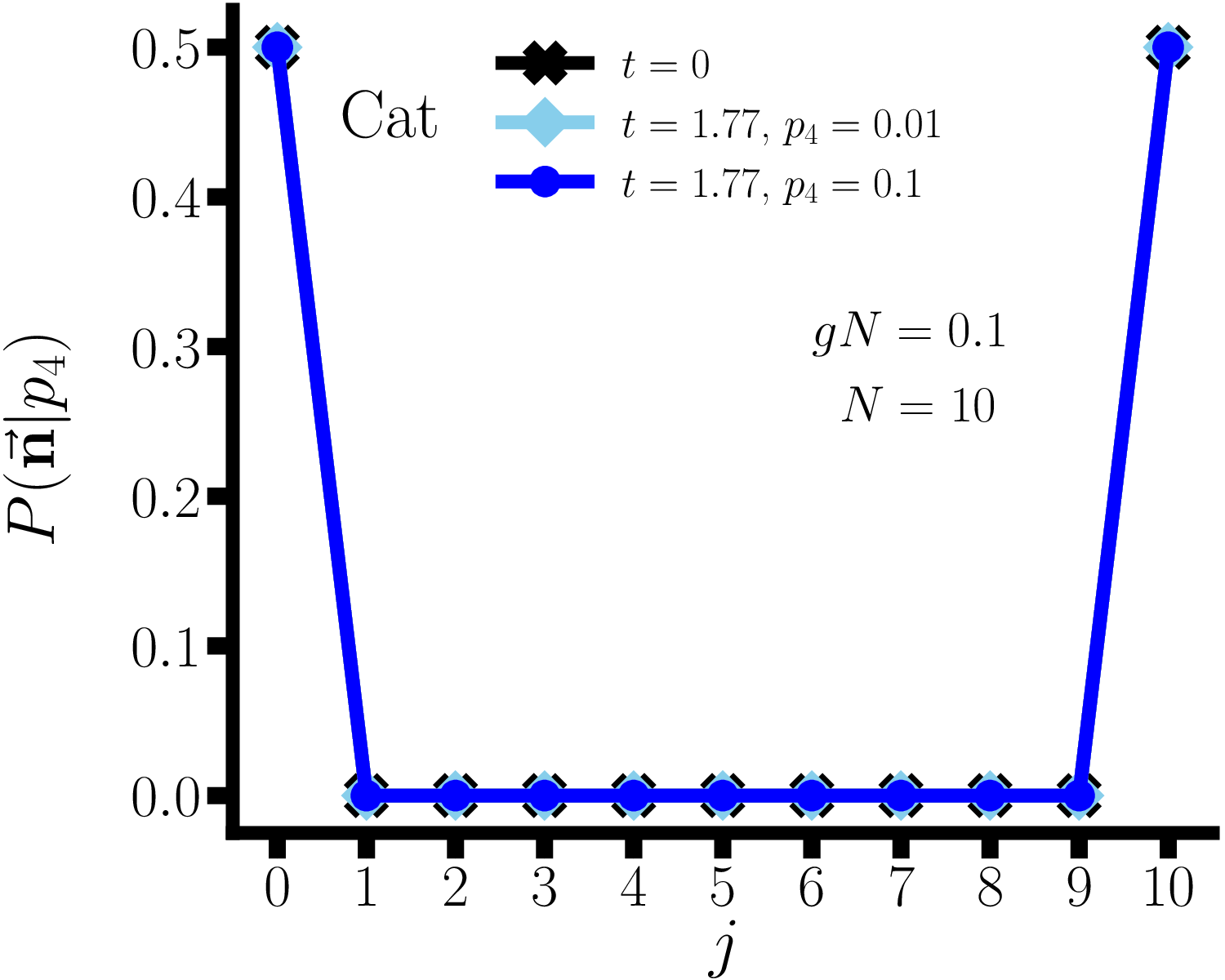}
\includegraphics[width=0.4\columnwidth]{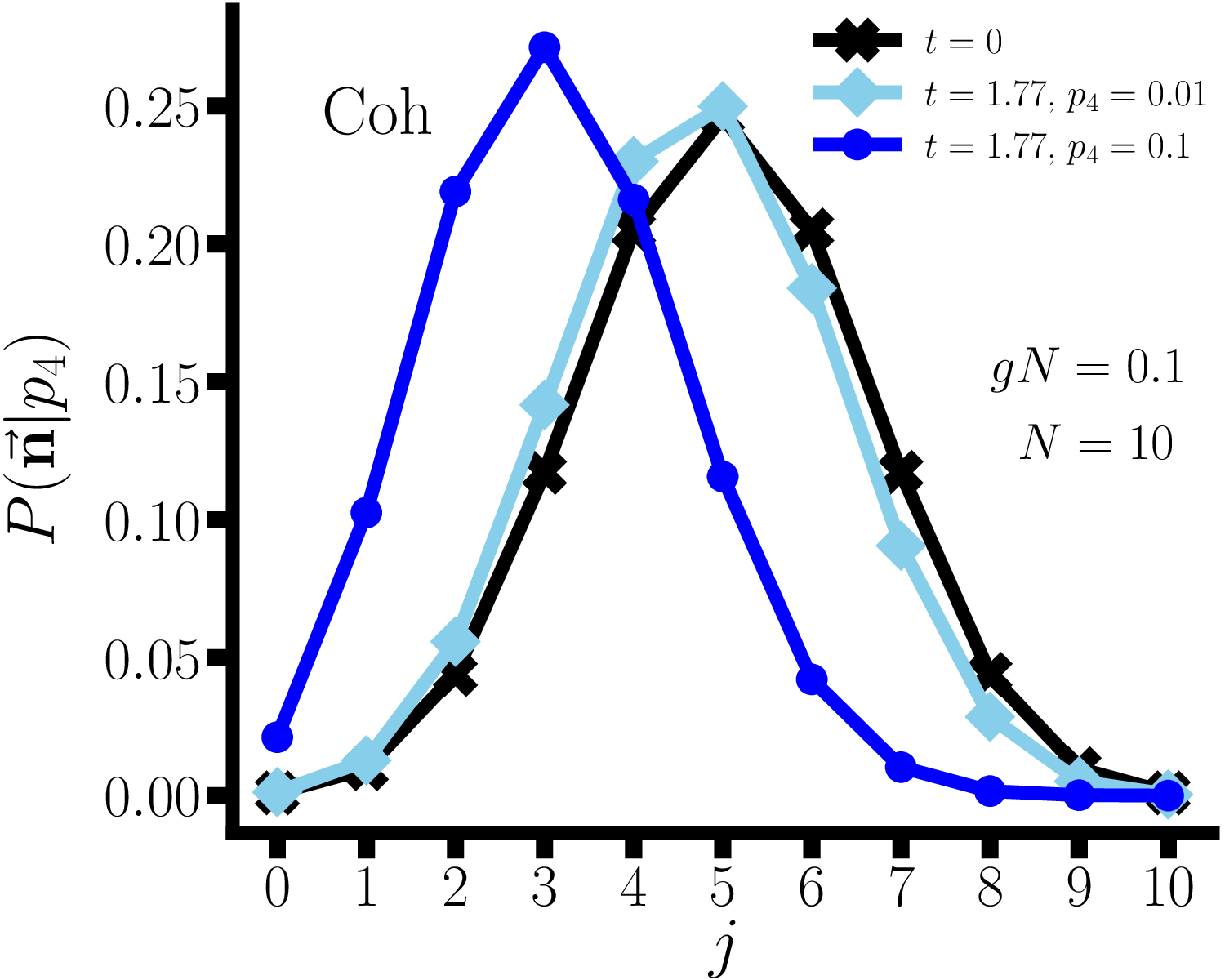}
\caption{\label{fig:scpd}
Probability distributions for the measurement outcome $\mathbf{n}=(n_\text{L},n_\text{R})=(N-j,j)$
with self-consistency taken into account.}
\end{figure}

For further illustration of the importance of self-consistency, see Fig.~\ref{fig:scpd}.
These plots show how the probability distribution, calculated by Eq.~\eqref{eq:scprob},
changes according to the  double-well  metrological scenario of the main text.
For the cat state (left), delocalization of orbitals is weak, and the probability distribution almost does 
not change even within the self-consistent framework.
Then self-consistent metrology simply repeats the results of conventional two-mode interferometry.
However, in the case of a spin-coherent state (right), the probability distribution evolves in time because of delocalizing orbitals.
The conventional two-mode interferometry, i.e., Eq.~\eqref{NSC_P}, predicts that the probability distribution remains identical to the initial one at $t=0$ (black crosses).
On the other hand, self-consisten metrology, i.e., Eq.~\eqref{SC_P}, predicts that the likelihood gets biased towards the left, and this is accurate since the particles are mostly located at the left well when $p_4>0$.
As $p_4$ increases, the likelihood gets even more biased,
and this dependence on $p_4$ results in a nonvanishing CFI.

\subsubsection{Further results on MLE statistics}
Additional details on the MLE are supplied in Fig.~\ref{fig:mle-supplementary}. % on the following page.
The top left shows the mean of the maximum likelihood estimator
with respect to the final state that has evolved under the true value of $p_4$.
To calculate the mean, the probability distribution first needs to be composed.
%; see section %Appendix \ref{sec:mctdh-probdist}.
One measurement outcome is used at a single time of estimation, i.e., $\nu=1$.
When the true value of $p_4$ is $0.1$, $\langle X_\text{est}\rangle_X\simeq0.0926$.
The top right plot shows the mean of the maximum likelihood estimator versus $\nu$ when $p_4=0.1$,
which is the number of measurement outcomes for a single estimation of $p_4$.
As $\nu$ increases, $\langle X_\text{est}\rangle_X$ converges at around $0.11$,
which implies a bias of MLE.
The bottom left shows this bias of the MLE,
where $\big|\partial\langle X_\text{est}\rangle_X/\partial X\big|\simeq1$ near $p_4\simeq0$,
which however does not hold as $p_4$ deviates more significantly from zero.
Finally, the bottom right shows the ratio between the mean-square deviation and the Cram\'er-Rao lower bound (CRLB).
This ratio decreases as $\nu$ increases: 
The MLE is known to make the mean-square deviation $\langle(\delta X_\text{est})^2\rangle_X$ converge to the CRLB
as $\nu\rightarrow\infty$ according to the central limit theorem, which is thus confirmed. % herewith.
\medskip

\begin{figure}[h]
\includegraphics[width=0.33\columnwidth]{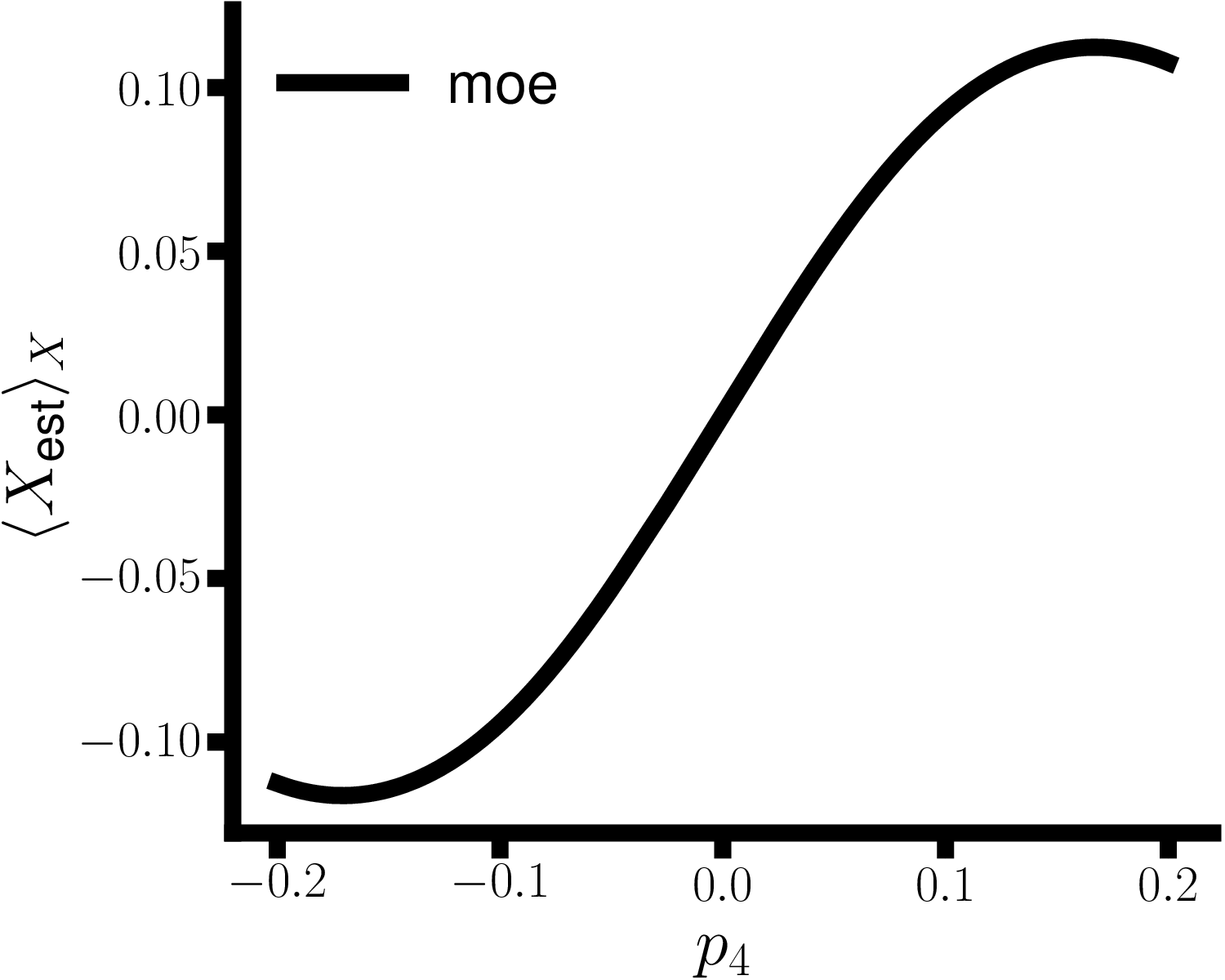}
\includegraphics[width=0.33\columnwidth]{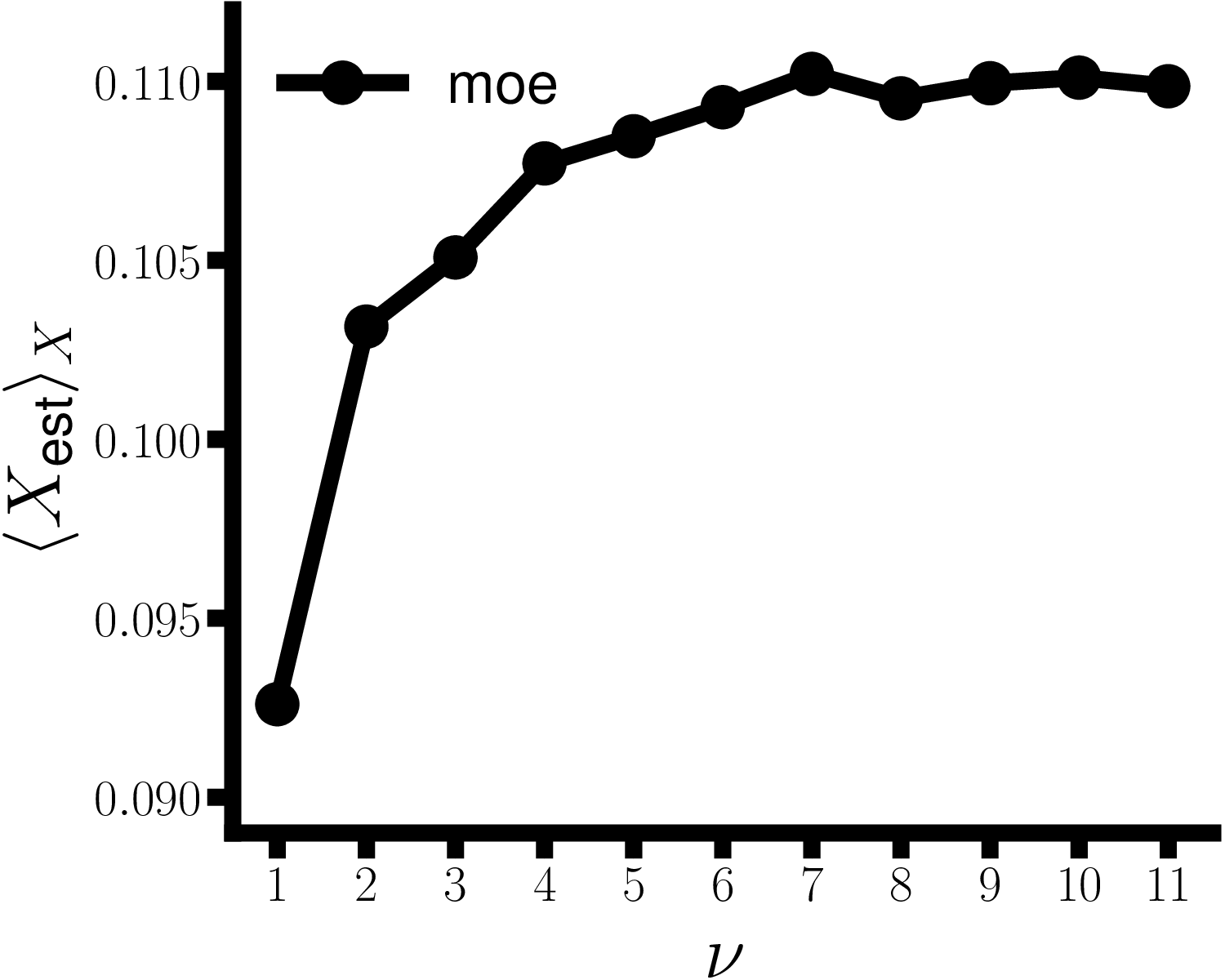}
\includegraphics[width=0.33\columnwidth]{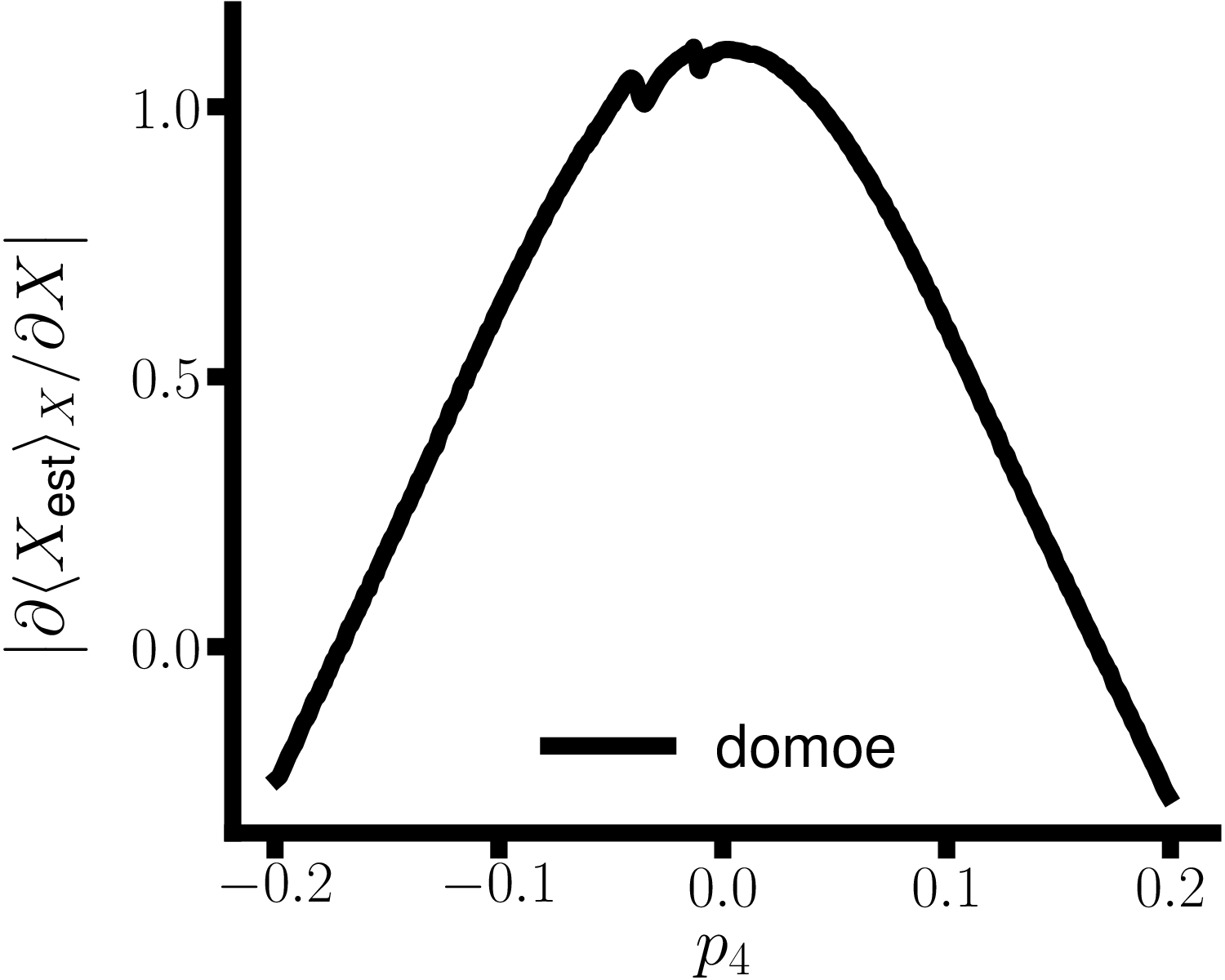}
\includegraphics[width=0.33\columnwidth]{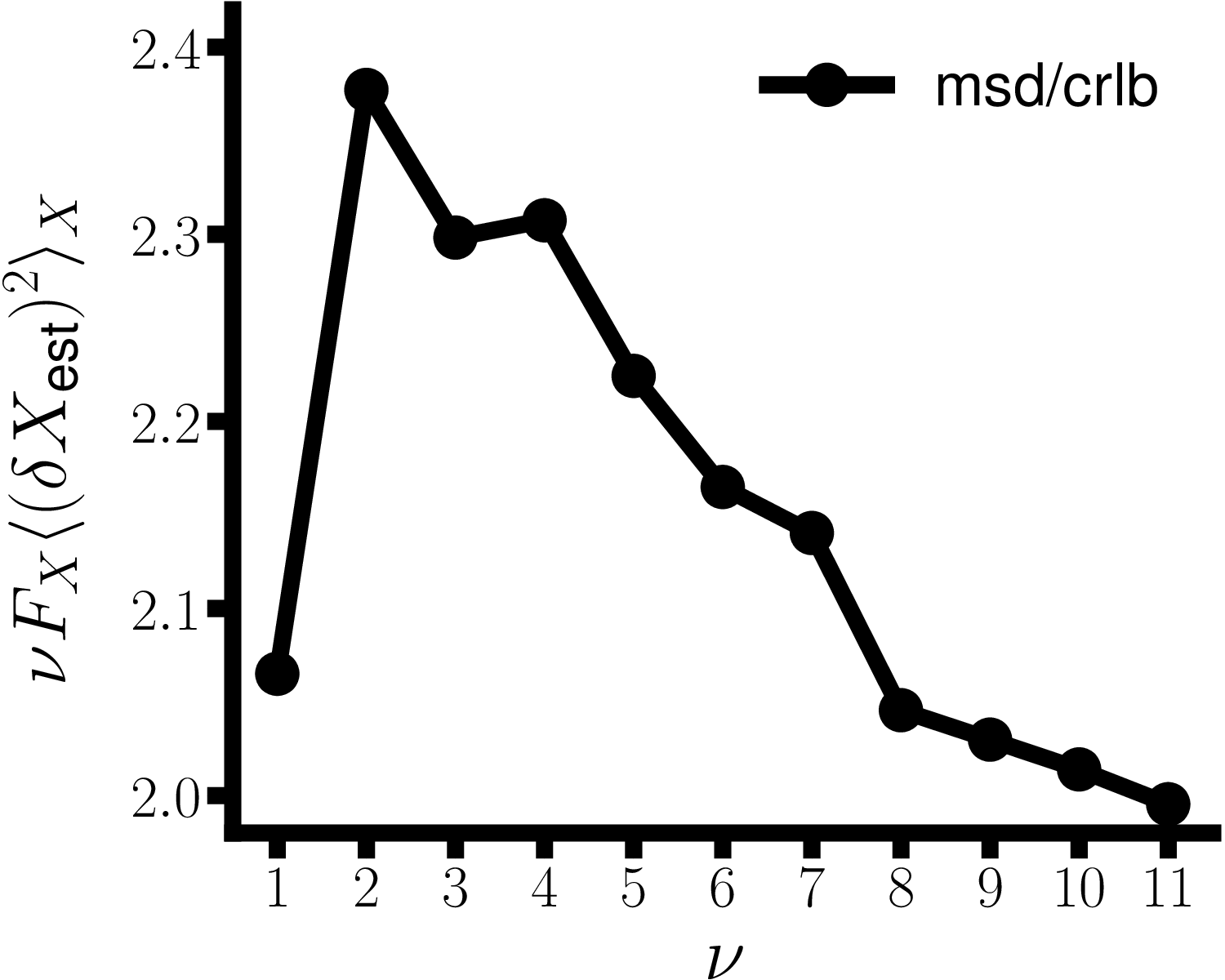}
\caption{\label{fig:mle-supplementary} Additional results on maximum likelihood estimation, 
where ``moe'' is the mean of estimates ($\langle X_\text{est}\rangle_X$),
``msd'' is the mean-square deviation ($\langle(\delta X_\text{est})^2\rangle_X$),
and ``domoe'' stands for the absolute value of the derivative of the mean of estimates ($\big|\partial\langle X_\text{est}\rangle_X/\partial X\big|$).}
\end{figure}

\subsection{V. Convergence of results for increasing number of modes: $M=2$, $M=3$, and $M=4$}

In the main text, it is shown that even for a weakly interacting gas,
studying the CFI can exhibit a discrepancy between two-mode interferometry and self-consistent metrology.
By monitoring $\rho_\text{tm}$, the validity of the two-mode approximation, $M=2$, had been confirmed in the time interval of interest, and accordingly, two modes are used in the self-consistent framework, too.
In the weakly interacting regime, two modes explain the primary quantum dynamics of the system
and thus, the quantum Fisher information which is governed by the final state can be predicted 
suffiiciently accurate by two-mode metrology.
Since interparticle interaction is weak, i.e., $gN=0.1$,
$M=2$ is thus used for the self-consistent results. 

\begin{figure}[b]
\includegraphics[width=0.33\columnwidth]{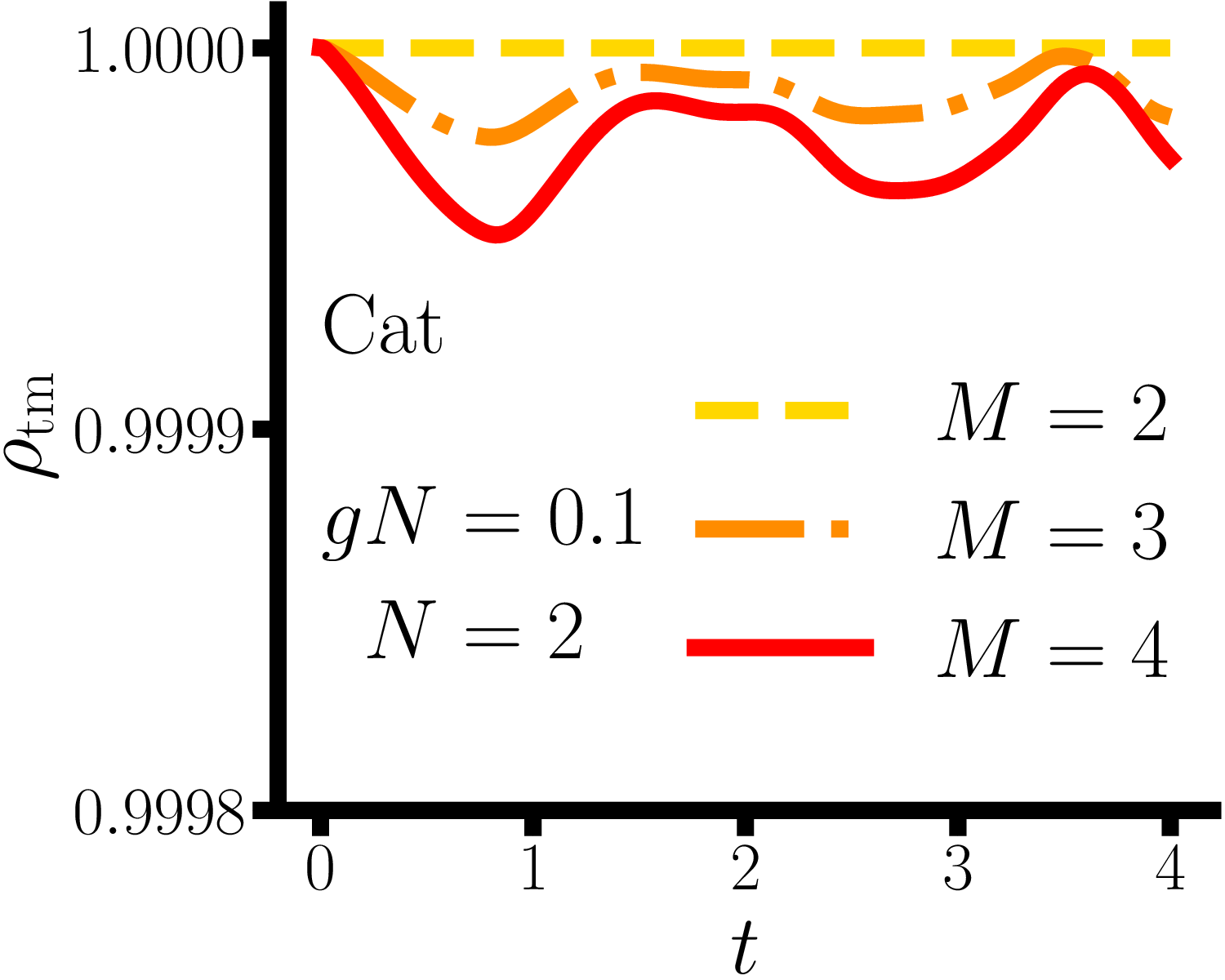}
\includegraphics[width=0.33\columnwidth]{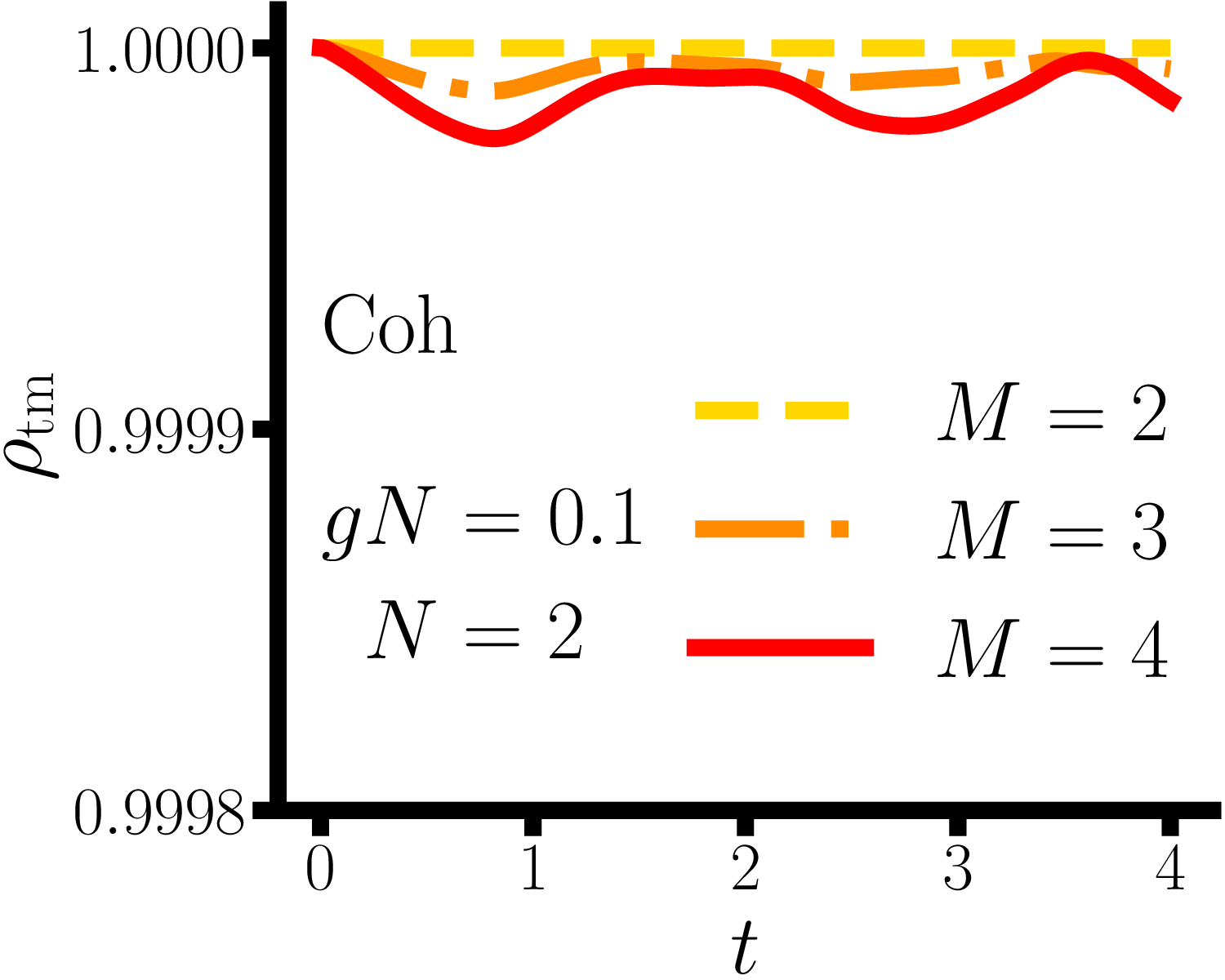}
\caption{\label{fig:two-mode2} Monitoring the two-mode truncation after $p_4$ is turned on,
by verifying whether $\rho_{\rm tm}= (\rho_1+\rho_2)/N\lesssim 1$ (we put $N=10$), with 
$gN$ values as indicated. Cat state is on the left and spin-coherent state on the right.}
\end{figure}

%, which displays good agreement of QFI with those of two-mode interferometry.}

%The full power of self-consistent metrology is the ability to calculate metrological quantities
%in regimes of any interaction strength by introducing additional orbitals.
To verify the validity of the two-mode approximation, we compare here with the $M=3$ and $M=4$ cases %are tested in the
in a similar regimes of parameters, namely, $gN=0.1$, $N=2$, and $t=0\sim 4$. 
The validity of two-mode approximation is checked in Fig.~\ref{fig:two-mode2}.
For $M=3$ and $M=4$, it is evident that $\rho_\text{tm}\lesssim1$, implying that the two-mode approximation works; the additional modes are rarely occupied during this time period. The impact of the additional modes is thus not crucial.

\begin{figure}[b]
\includegraphics[width=0.33\columnwidth]{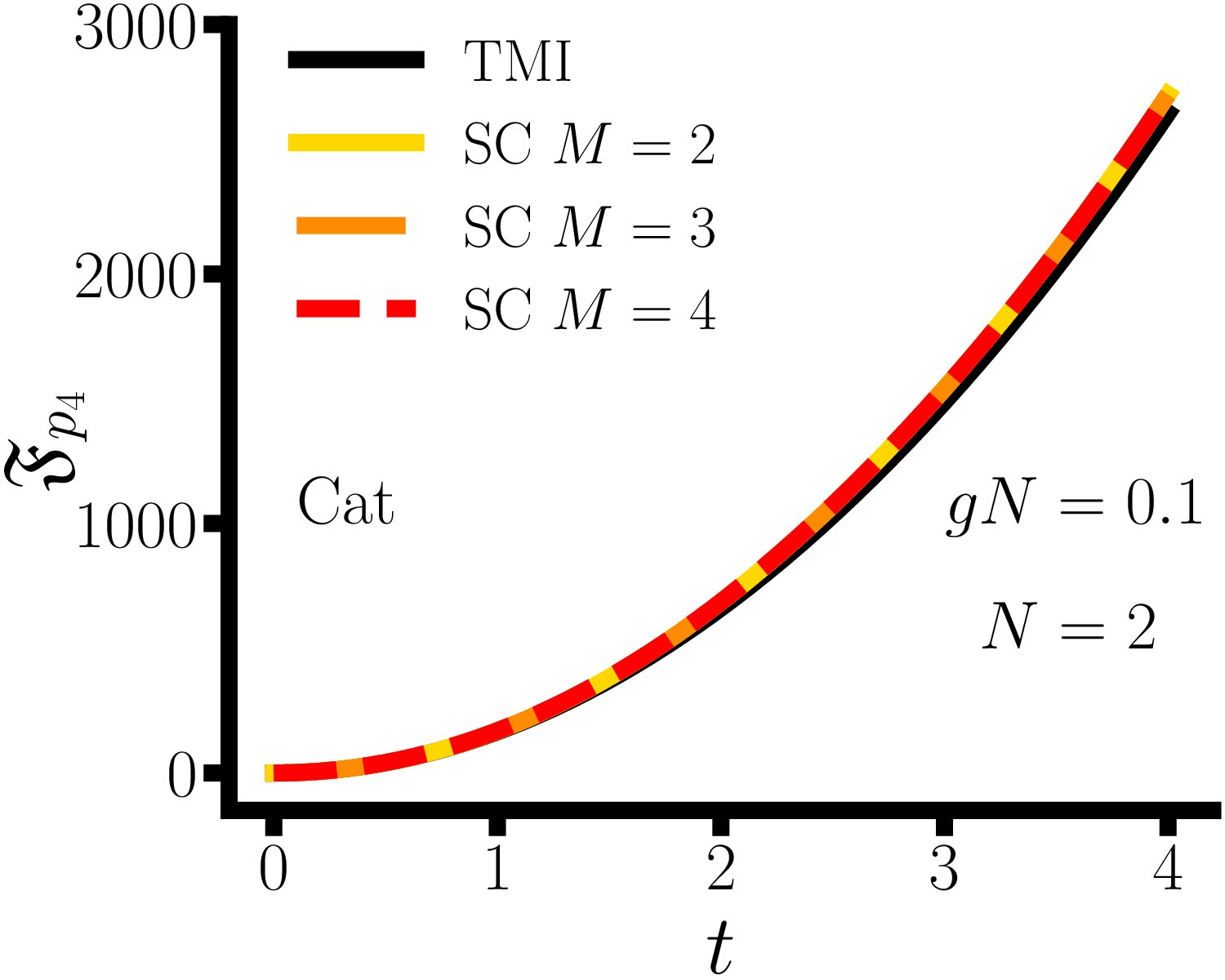}
\includegraphics[width=0.33\columnwidth]{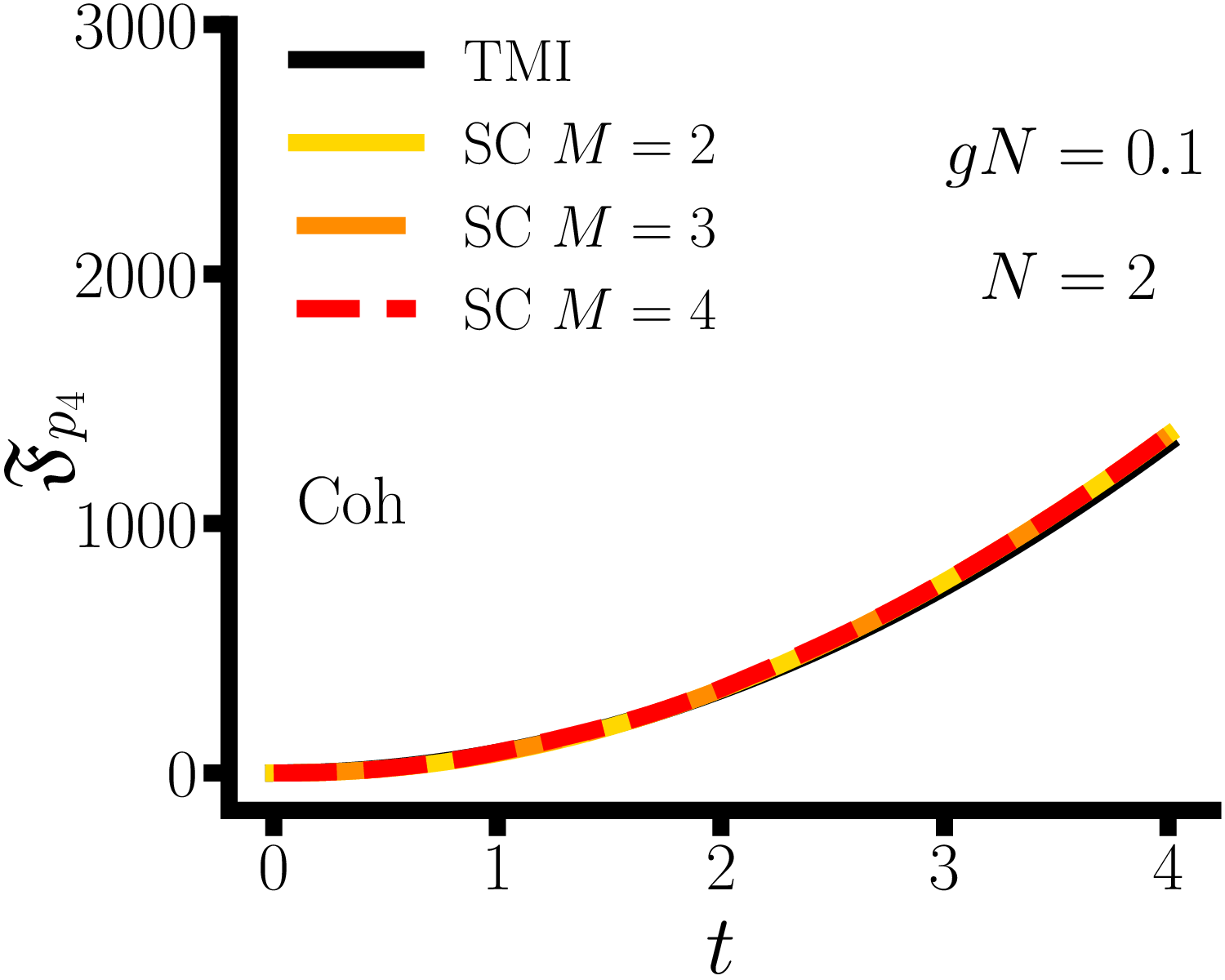}
\includegraphics[width=0.33\columnwidth]{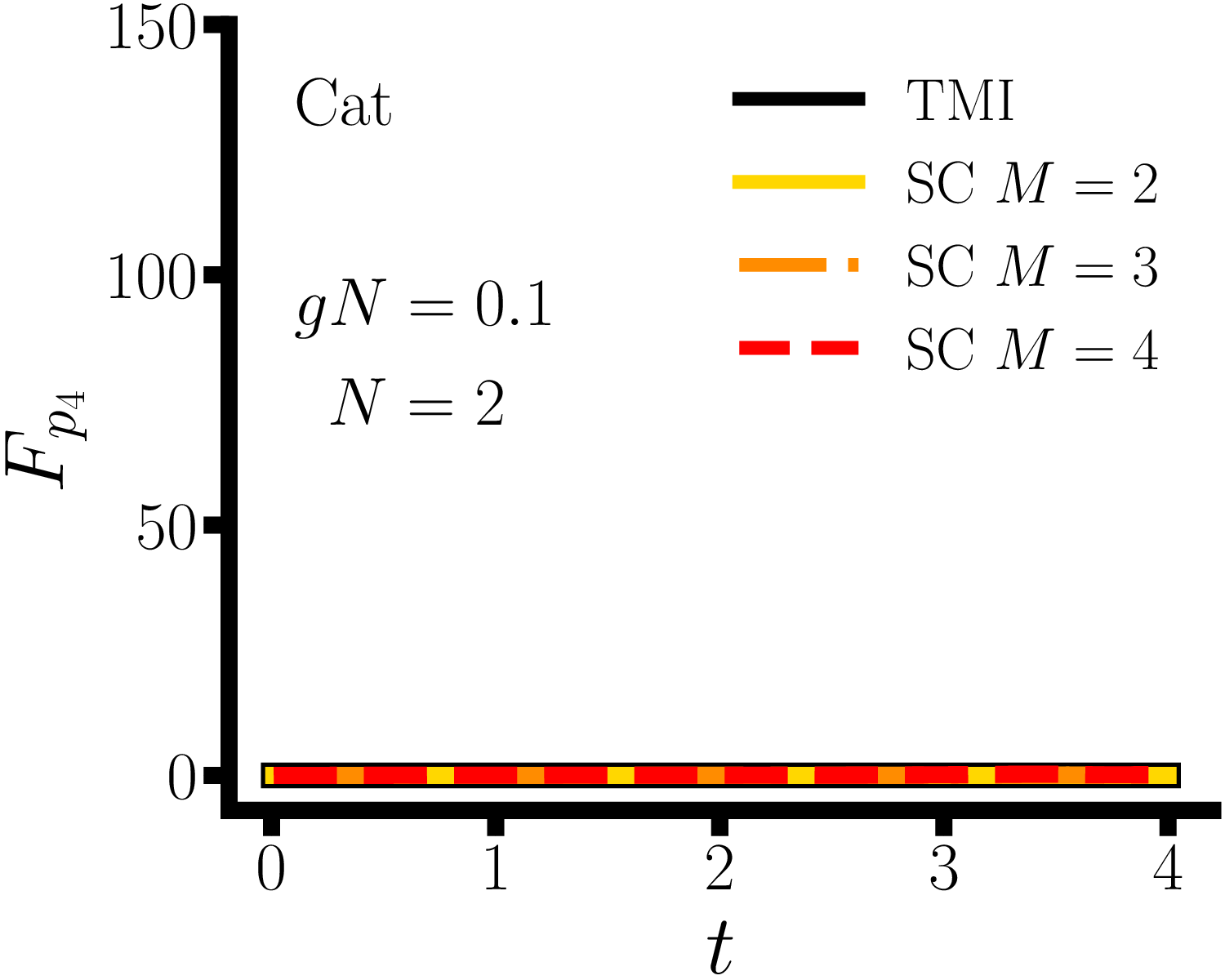}
\includegraphics[width=0.33\columnwidth]{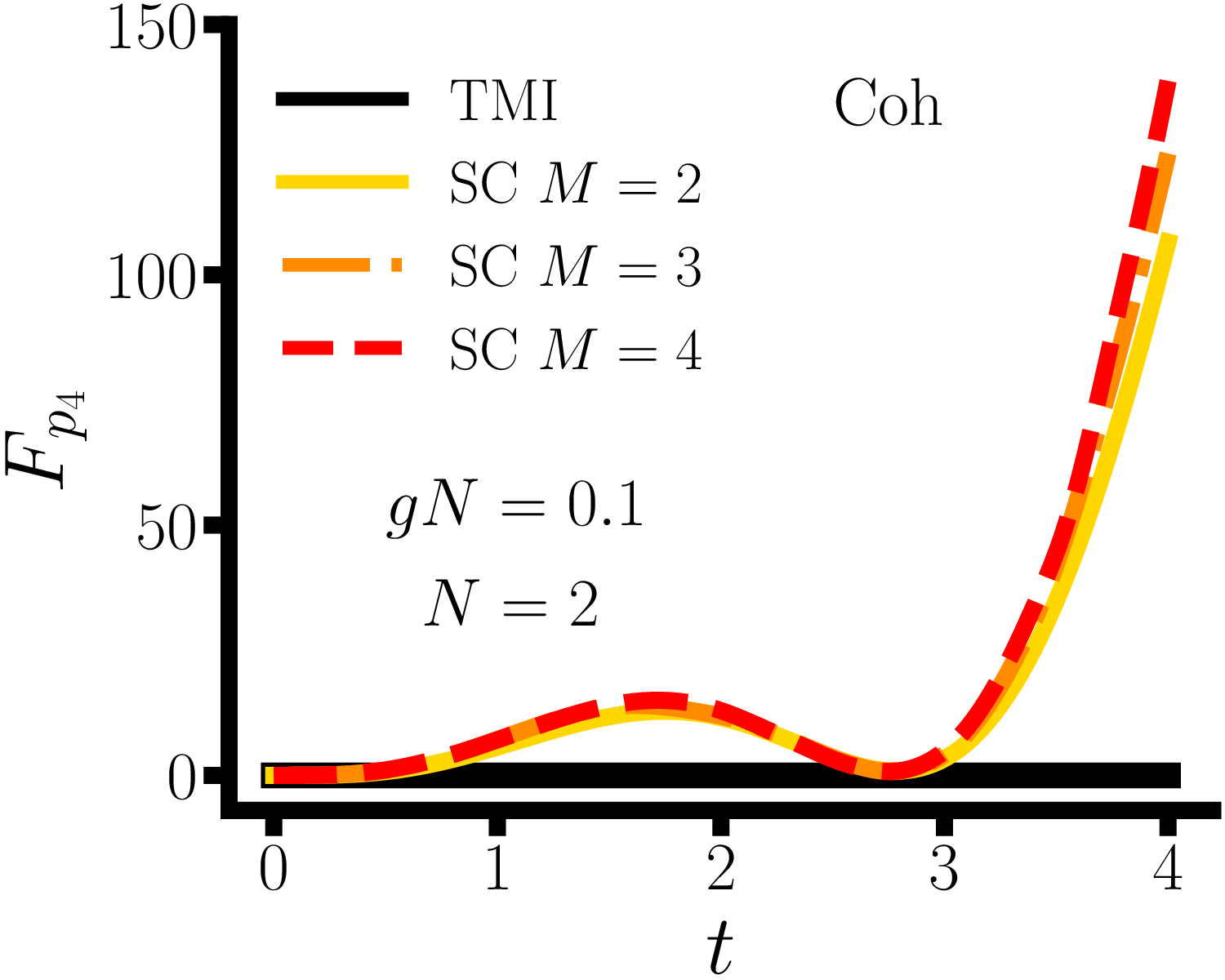}
\caption{\label{fig:qfi2} First and second row display quantum ($\mathfrak{F}$) and classical (F) Fisher information, respectively, plotted versus time $t$, for cat and coherent (coh) states, respectively.}
\end{figure}

This also can be seen in the QFI with respect to the parameter $p_4$; see upper row in Fig.~\ref{fig:qfi2}.
For the weakly interacting gas, a small number of modes, e.g. here $M=2$, 
is sufficient to describe the dynamics of the system and thus non-self-consistent two-mode interferometry calculates the QFI rather precisely.
The self-consistent results with two (yellow solid), three (orange dash-dotted), and four (red dotted) modes are
almost identical, and are not significantly different from the value of two-mode interferometry (black solid).

We have seen in the main text that the CFI for $p_4$ can significantly differ when evaluated in the non-self-consistent and self-consistent frameworks.
The lower row in Fig.~\ref{fig:qfi2} shows that the inclusion of more orbitals reproduces (and thus also further 
emphasizes) the different predictions made for $M=2$ in the main text,  when comparing the CFI in a self-consistent metrological framework relative to that in a non-self-consistent one. In the bottom left of Fig.~\ref{fig:qfi2}, with a cat distribution of the coefficients of the initial state, the CFI for a certain measurement, i.e., measuring the number of particles in the left and right well,
remains almost zero.
On the other hand, in the bottom right of Fig.~\ref{fig:qfi2}, with a spin-coherent distribution of the 
coefficients of the initial state, the CFI is nonvanishing as time passes by.
This is confirmed for $M=2$ (yellow solid), $M=3$ (orange dash-dot), and $M=4$ (red dotted).
The nonvanishing CFI {\em cannot} be predicted in the conventional framework of non-self-consistent two-mode interferometry.  The self-consistent approach thus makes it possible at all to accurately evaluate the precision limit for estimating the parameter $p_4$.

\end{widetext}
% The \nocite command causes all entries in a bibliography to be printed out
% whether or not they are actually referenced in the text. This is appropriate
% for the sample file to show the different styles of references, but authors
% most likely will not want to use it.
%\nocite{*}

% Create the reference section using BibTeX:

\end{document}